\documentclass[prb,twocolumn,showpacs,amsmath,amssymb,superscriptaddress]{revtex4-2}
\usepackage{xcolor}
\usepackage[pdftex]{graphicx,hyperref}
\hypersetup{colorlinks = true, urlcolor = blue, linkcolor = blue, citecolor = blue}
\usepackage{enumerate}
\usepackage{float}

\newcommand{\nc}{\newcommand}
\nc{\be}{\begin{equation}} \nc{\ee}{\end{equation}}
\nc{\bea}{\begin{eqnarray}} \nc{\eea}{\end{eqnarray}}
\nc{\bean}{\begin{eqnarray*}} \nc{\eean}{\end{eqnarray*}}
\nc{\dg}{\dagger}
\nc{\ua}{\uparrow} \nc{\da}{\downarrow}
\nc{\lag}{\langle} \nc{\rag}{\rangle}

\begin{document}
	
	\title{Higher-Order Topological Dirac Superconductors}
	\author{Rui-Xing Zhang}
	\email{ruixing@umd.edu}
	\author{Yi-Ting Hsu}
	\author{S. Das Sarma}
	\affiliation{Condensed Matter Theory Center and Joint Quantum Institute, Department of Physics, University of Maryland, College Park, Maryland 20742-4111, USA}
	
	\begin{abstract}
		We introduce higher-order topological Dirac superconductor (HOTDSC) as a new gapless topological phase of matter in three dimensions, which extends the notion of Dirac phase to a higher-order topological version. Topologically distinct from the traditional topological superconductors and known Dirac superconductors, a HOTDSC features Majorana hinge modes between adjacent surfaces, which are direct consequences of the symmetry-protected higher-order band topology manifesting in the system. Specifically, we show that rotational, spatial inversion, and time-reversal symmetries together protect the coexistence of bulk Dirac nodes and hinge Majorana modes in a seamless way. We define a set of topological indices that fully characterizes the HOTDSC. We further show that a practical way to realize the HOTDSC phase is to introduce unconventional odd-parity pairing to a three-dimensional Dirac semimetal while preserving the necessary symmetries. As a concrete demonstration of our idea, we construct a corresponding minimal lattice model for HOTDSC obeying the symmetry constraints. Our model exhibits the expected topological invariants in the bulk and the defining spectroscopic features on an open geometry, as we explicitly verify both analytically and numerically. Remarkably, the HOTDSC phase offers an example of a “higher-order” topological quantum critical point, which enables realizations of various higher-order topological phases under different symmetry-breaking patterns. In particular, by breaking the inversion symmetry of a HOTDSC, we arrive at a higher-order Weyl superconductor, which is yet another new gapless topological state that exhibits hybrid higher-order topology. 
		
	\end{abstract}
	\date{\today}
	\maketitle
	
\section{Introduction}
Dirac physics is one of the central concepts driving the intellectual revolution of topological phases in condensed matter physics \cite{hasan2010rmp,qi2011rmp}. Since the seminal proposal of quantum spin Hall effect in graphene with a hypothetically large spin-orbit coupling \cite{kane2005quantum,kane2005z2}, it has been known that gapping a Dirac system is a natural way to achieve a gapped topological state. On the other hand, boundary modes with Dirac dispersion often emerge on the ($d-1$)-dimensional boundaries of a $d$-dimensional topological phase, enforced by the bulk-boundary correspondence principle. For example,
two-dimensional (2d) and three-dimensional (3d) time-reversal-invariant topological insulators (TI) are known to host one-dimensional (1d) helical Dirac edge states \cite{kane2005quantum,kane2005z2,bernevig2006quantum,konig2007quantum} and 2d Dirac surface states \cite{hasan2010rmp,qi2011rmp}, respectively. A 3d massless Dirac fermion is by definition four-fold-degenerate in its energy dispersion, which can only be stabilized in solids in the presence of certain crystalline symmetries \cite{wang2012dirac,yang2014classification,young2017filling,wieder2018wallpaper}. A bulk electronic system with such 3d massless Dirac fermion is known as a 3d Dirac semimetal, which was first observed with angle-resolved photoemission spectroscopy in Na$_3$Bi \cite{wang2012dirac} and Cd$_3$As$_2$ \cite{wang2013three}. In those materials, the bulk Dirac points are supported by the combined protection of time-reversal symmetry (TRS), spatial inversion symmetry, and certain out-of-plane rotational symmetries. When projected onto the surface, the bulk Dirac points are connected by arc-like surface states on the Fermi surface, which manisfests the topological nature of a DSM. Such DSMs can be driven to a 3d TI or a Weyl semimetal by explicit or spontaneous symmetry breakings. Another intriguing 3d Dirac phase is the 3d Dirac superconductor (DSC) \cite{yang2014dirac}, whose defining features are gapless Dirac points in the bulk Bogoliubov-de Gennes (BdG) spectrum and Majorana Fermi-arc surface states. Candidate materials for 3d DSCs include Cu$_x$Bi$_2$Se$_3$ \cite{yang2014dirac}, doped DSMs \cite{kobayashi2015topological,hashimoto2016superconductivity}, and iron-based superconductors \cite{zhang2019multiple}.

\begin{figure*}[t]
	\centering
	\includegraphics[width=0.95\textwidth]{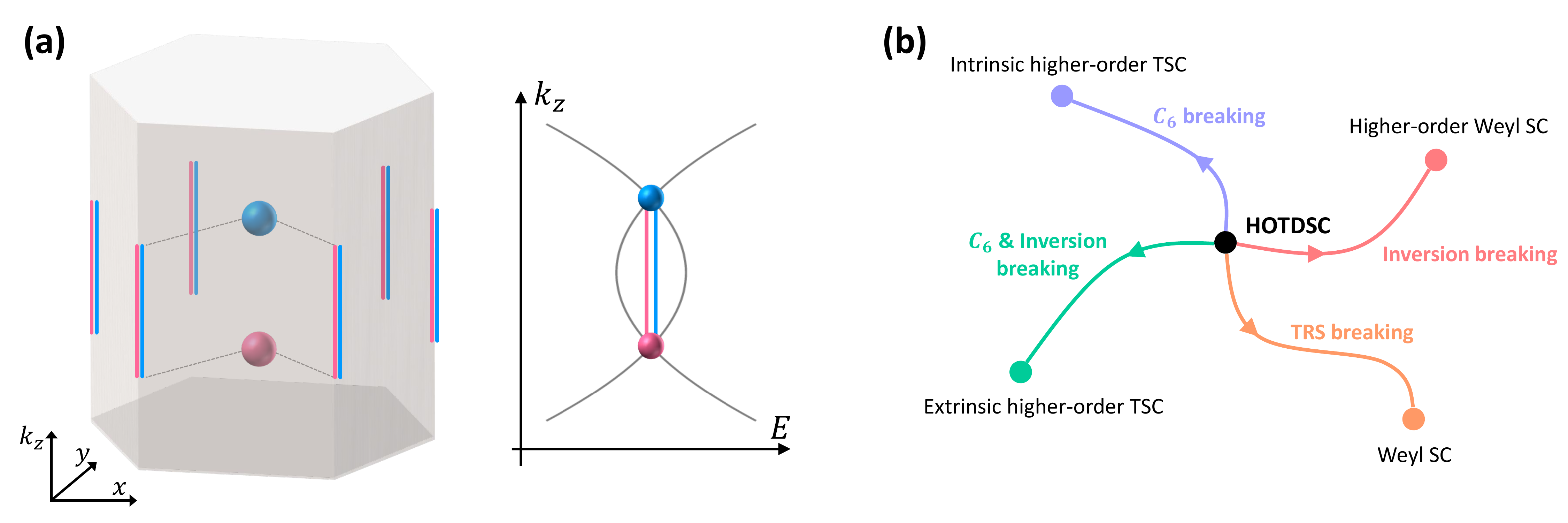}
	\caption{In (a), we plot a schematic of a higher-order topological Dirac superconductor in a hexagonal prism geometry. This exotic phase is featured by the coexistence of BdG Dirac nodes in the 3d bulk (the colored spheres) and Majorana modes on the 1d hinges (the red and blue lines). We show a schematic dispersion plot for the HOTDSC phase in the prism geometry, which describes how the hinge Majorana modes connect the bulk Dirac nodes. In (b), we demonstrate the higher-order topological Dirac superconductor as a higher-order topological quantum critical point. Various topological or higher-order topological phases can be achieved upon explicit or spontaneous symmetry breaking. A detailed discussion is presented in Sec. \ref{Sec: QCP}.}
	\label{Fig: Schematic}
\end{figure*}

On the other hand, a recent development in the topological classification of matter is the extension of band topology into a ``higher-order" version \cite{benalcazar2017quantized,zhang2013surface,benalcazar2017electric,song2017d-2,schindler2018higher,langbehn2017reflection,khalaf2018higher,slager2015impurity,trifunovic2019higher,trifunovic2020higher},
where the topologically protected boundary modes can live in a lower codimension than those in traditional topological materials. 
Specifically, we adapt the definition in which an $\Omega$-th order topological state has anomalous gapless modes on its $d-\Omega$ dimensional boundary. In this definition, $\Omega=1$ and $\Omega>1$ correspond respectively to traditional and higher-order topological phases. 
For example, a 3d second-order TI has energy gaps on its 2d surfaces, but the hinges connecting different surfaces can host 1d channels that penetrate both bulk and surface gaps. 
The robustness of this new type of higher-order topology often originates from the protection of certain crystalline symmetries in the sense that the boundary modes of a higher-order topological phase cannot be removed without either closing the bulk gap or breaking these symmetries \cite{khalaf2018symmetry}.
This type of higher-order topological phases are often referred to as ``intrinsic", which is in contrast to the ``extrinsic" ones that lack crystalline symmetry protection.

While higher-order TIs have attracted considerable attention, there has been relatively little research on higher-order topology in superconductors \cite{shapourian2018topological,wang2018weak,wang2018high,yan2018majorana,liu2018majorana,ono2018unified,zhu2018tunable,pan2018lattice,volpez2019second,zhu2019second,ghorashi2019second,bultinck2019three,wu2019higher,peng2019proximity}. 
One major difference between the two cases is that in the latter, the hinge or corner-localized states are essentially Majorana modes.
So far, extrinsic higher-order topological superconductivity has been theoretically proposed in TI/unconventional superconductor heterostructures \cite{wang2018weak,wang2018high,yan2018majorana,liu2018majorana}, iron-based superconductors \cite{zhang2019helical,zhang2019higher,wu2019high}, and other platforms \cite{pan2018lattice,peng2019proximity,wu2019higher,volpez2019second}, whereas inversion-protected intrinsic superconductivity has been proposed in gated monolayer WTe$_2$ \cite{hsu2019inversion} and doped ferromagnetic nodal semimetals \cite{ahn2019higher}. 
Overall, intrinsic higher-order 3d topological superconductors (TSC) remain largely unexplored.  

Here we raise the following important conceptual questions: {\it  Can an intrinsic 3d higher-order version of superconducting Dirac phases in principle exist? If so, what are the protecting symmetries?} In spite of several proposals on electronic higher-order topological semimetallic phases \cite{lin2018topological,cualuguaru2019higher,szabo2019strain,wang2018higher,ahn2019higher}, we are not aware of any literature on exploring the coexistence of higher-order topology and nodal superconductivity, which still remains an important open question.

In this work, we introduce a new type of gapless topological phase, the {\it higher-order topological Dirac superconductor} (HOTDSC), as the first example of a symmetry-protected higher-order topological nodal superconducting phase \cite{fnote2}.
The HOTDSC gets its name from simultaneously hosting 3d bulk Dirac nodes, 2d gapped surface states, and 1d flat-band-like hinge Majorana modes in the BdG spectrum, all of which are enforced by symmetries [see Fig. \ref{Fig: Schematic} (a)]. 
We first show that the bulk Dirac points and the Majorana hinge modes are protected by $C_6$ rotation and inversion symmetry respectively. We then define a set of topological invariants accordingly that fully characterize this phase. Importantly, we point out that introducing superconductivity to doped DSMs can be a practical way to experimentally realize this exotic higher-order phase.
Through a systematic study of superconducting DSMs with various rotational symmetries [see Table \ref{Table: From DSM to DSC}], we find that $C_6$ is the only rotation symmetry that can protect our HOTDSC. 

To develop the theory, we construct a minimal model for the HOTDSC phase by introducing a time-reversal pairing gap that is odd under both $C_6$ rotation and inversion to a $C_6$-symmetric doped DSM.
Specifically, we show analytically and numerically that our minimal model, both in the continuum limit and on a hexagonal prism lattice, not only exhibits the expected topological invariants, but also possesses a pair of robust bulk Dirac points, gapped 2d surfaces, and Majorana hinge modes. Hence the bulk-boundary correspondence for the HOTDSC phase is explicitly demonstrated in our model.

Importantly, our proposed HOTDSC phase also offers an example of a {\it higher-order topological quantum critical point}. As shown in Fig. \ref{Fig: Schematic} (b), a variety of higher-order topological phases could be achieved upon different symmetry breakings of a HOTDSC, including an intrinsic/extrinsic higher-order TSC, a Weyl superconductor and a {\it higher-order Weyl superconductor}. In particular, the higher-order Weyl superconductor is a qualititatively new type of gapless topological state that has never before been defined or studied. Different from a HOTDSC, a higher-order Weyl superconductor features not only 3d bulk Weyl nodes connected by 2d surface Majorana Fermi arcs in the BdG spectrum, but also possesses coexisting 1d flat-band-like hinge Majorana modes. Remarkably, the higher-order Weyl superconductor offers a natural example for ``hybrid higher-order topology" \cite{bultinck2019three}, where 2d surface state and 1d hinge modes coexist and are protected against mixing with each other by the translational symmetry. 

The paper is organized as follows. In Sec. \ref{Sec: Symmetry indicators}, we present a detailed discussion on the bulk topological invariants that are crucial for defining the HOTDSC phase. By clarifying the topological nature of superconducting DSMs, we identify the symmetry requirements for HOTDSC phase. 
In Sec. \ref{Subsec: Symmetry for BdG Systems}, we briefly review the extension of symmetry operations from a normal electron system to its BdG counterpart. In Sec. \ref{Subsec: topological charge for DSMs}, we first review the definition of topological charge for electronic DSMs and further generalize the theory to describe BdG Dirac systems. This extension allows us to establish the deep connection between superconducting DSMs and Dirac superconductors. In Sec. \ref{Subsec: inversion symmetry indicator}, we discuss how to use a symmetry indicator $\kappa_{2d}$ for two-dimensional inversion-symmetric BdG systems to diagnose Majorana hinge modes in three-dimensional HOTDSC. In Sec. \ref{Subsec: mirror Chern number}, we develop a simple relation between the mirror Chern number of a BdG system and that of its normal part. The above discussions on the topological indices pave the way for clarifying the required conditions of HOTDSC, which is concluded in Sec. \ref{Subsec: topological requirements for HOTDSC}.

In Sec. \ref{Sec: Model}, we present a minimal lattice model for the HOTDSC phase and establish the defining properties of the HOTDSC phase both analytically and numerically. 
In Sec. \ref{Subsec: lattice minimal model}, we start by introducing a tight-binding model for 3d $C_6$-symmetric DSM on a hexagonal lattice. In Sec. \ref{Subsec: Odd parity pairing}, we classify pairing terms that satisfy the symmetry requirement for HOTDSC, thus leading to the non-trivial bulk Dirac physics. In Sec. \ref{Subsec: Effective boundary theory}, we analytically solve our superconducting model for the low-energy surface state in a cylinder geometry and construct an effective boundary BdG theory for our model. This effective theory shows an anisotropic surface pairing gap that directly implies the existence of hinge Majorana states. In Sec. \ref{Subsec: hinge Majorana}, we numerically calculate the energy spectrum in an infinite long hexagonal prism geometry to unambiguously and explicitly demonstrate the co-existence of bulk Dirac physics and hinge Majorana physics.

In Sec. \ref{Sec: QCP}, we establish the HOTDSC phase as a higher-order topological quantum critical point and introduce the concept of higher-order Weyl SC. We then show how various topological phases (especially the higher-order Weyl physics) naturally emerge from the HOTDSC by breaking different symmetries. 
Finally in Sec. \ref{Sec: Conclusion}, we summarize our results and discuss possible directions for the experimental realization of our predictions. 

\section{Symmetries and Topological Invariants for HOTDSC}
\label{Sec: Symmetry indicators}

In this section, we establish the theoretical framework for a HOTDSC phase by introducing the crucial symmetries and the corresponding topological invariants. Specifically, we first define a topological charge $Q_j$ for $C_n$ rotation symmetry and a mirror Chern number ${\cal C}_M$ for a mirror symmetry in z direction. We then apply an inversion symmetry indicator $\kappa_{2d}$ for 2d time-reversal gapped superconductors to study a high-symmetry plane in a 3d HOTDSC. These three invariants govern the topological properties of the 3d bulk, 2d surfaces, and 1d hinges respectively in a 3d superconductor. Our recipe for realizing a HOTDSC are determined by both the symmetry constraints and the values of $\{Q_j,{\cal C}_M,\kappa_{2d}\}$.

\subsection{Symmetry for BdG Systems}
\label{Subsec: Symmetry for BdG Systems}
We start by reviewing the symmetry properties of a general BdG Hamiltonian 
\bea
H({\bf k}) = \begin{pmatrix}
	h({\bf k}) & \Delta({\bf k}) \\
	\Delta^{\dagger}({\bf k}) & -h(-{\bf k})^* \\
\end{pmatrix}
\eea
defined in the Nambu basis
\bea
\Psi ({\bf k}) = (c_{1,{\bf k}},c_{2,{\bf k}},...,c_{N,{\bf k}},c^{\dagger}_{1,-{\bf k}},c^{\dagger}_{2,-{\bf k}},...,c^{\dagger}_{N,-{\bf k}})^T,
\eea
where $c_{i,{\bf k}}$ ($c_{i,-{\bf k}}^{\dagger}$) annihilates an electron (a hole) at momentum \textbf{k} with band index $i=1,2,...,N$. $H({\bf k})$ satisfies the particle-hole symmetry (PHS)
\bea
\Xi = \begin{pmatrix}
	0 & I_N \\
	I_N & 0 \\
\end{pmatrix}{\cal K} = \tau_x {\cal K}.
\eea
$I_N$ is an $N\times N$ identity matrix and ${\cal K}$ is the complex conjugation. We define the Pauli matrix $\tau_i$ to characterize the particle-hole degree of freedom. 

Let us assume that the normal state Hamiltonian $h(k)$ is invariant under a unitary symmetry $\tilde{A}$ with $\tilde{A} h({\bf k}) \tilde{A}^{\dagger} = h(\tilde{A} {\bf k})$. When the pairing function satisfies 
\bea
\tilde{A} \Delta({\bf k}) \tilde{A}^T = \chi \Delta(\tilde{A}{\bf k}),
\label{Eq: Pairing transforms}
\eea
with $\chi$ being a $U(1)$ phase factor, the BdG Hamiltonian $H({\bf k})$ is invariant under the BdG extension of $\tilde{A}$, 
\bea
A = \begin{pmatrix}
	\tilde{A} & 0 \\
	0 & \chi \tilde{A}^*
\end{pmatrix}.
\label{Eq: BdG Unitary Symmetry}
\eea
When we further require $H(\textbf{k})$ to be invariant under the time-reversal operation $\Theta$, the requirement that $[\Theta,A]=0$ will impose a strong constraint on both $\chi$ and the pairing function. Specifically, the time-reversal operation for a general BdG system is given by
\bea
\Theta = \begin{pmatrix}
	T & 0 \\
	0 & T^* \\
\end{pmatrix}{\cal K},
\eea
where $T$ is the unitary part of the normal-state time-reversal operation. 
Together with Eq. \ref{Eq: BdG Unitary Symmetry}, we arrive at
\bea
\chi = \pm 1 \in \mathbb{R}.
\eea
Therefore, for a time-reversal-invariant BdG system that respects A symmetry, the pairing function can be either even ($\chi=1$) or odd $(\chi=-1)$ under $\tilde{A}$, following Eq. \ref{Eq: Pairing transforms}.

Next, we will consider time-reversal BdG Hamiltonians with symmetry $A$ being the rotational, mirror, and inversion symmetries, and discuss their corresponding topological indices.  
These three crystalline symmetries are necessary for protecting a HOTDSC phase, which guarantee gapless 3d Dirac nodes, gapped 2d surface states, and gapless 1d hinge modes, respectively.

\subsection{Rotation Topological Charge $Q_j$}
\label{Subsec: topological charge for DSMs}

In this subsection, we discuss how certain rotational symmetries can stabilize 3d bulk Dirac nodes in both Dirac semimetals (DSM) and Dirac superconductors (DSC), and discuss their corresponding topological indices.
Following Ref. \cite{kobayashi2015topological} and \cite{yang2015topological}, we first define a set of topological charges $\{\tilde{Q}_j\}$ for 3d DSM that (i) characterizes the existence of rotation-protected bulk Dirac nodes; (ii) only relies on the rotation eigenvalues at time-reversal-invariant momenta (TRIM) along the rotation axis (e.g. $\Gamma$ and $Z$). These topological charges are hence symmetry indicators. 
Based on the definition of $\{\tilde{Q}_j \}$, we will then define analogous topological charges $\{Q_j \}$ for Dirac superconductors (DSC). The deep connection between $\{\tilde{Q}_j \}$ and $\{Q_j \}$ offers a simple approach to determine what kind of DSC physics can be achieved by a given superconducting DSM system in the weak pairing limit. 

\subsubsection{Topological Charge for Dirac Semimetals}
\label{Subsubsec: DSM topo charge}

For a DSM whose bulk Dirac nodes are away from TRIMs, the presence of symmetry-protected bulk Dirac nodes can be diagnosed by a set of topological charges $\{\tilde{Q}_j\}$ defined on the rotation axis (e.g. $k_z$-axis) \cite{yang2015topological}.
Generally, for an electronic system with an $n$-fold rotation symmetry $\tilde{C}_n$, there are $n$ inequivalent irreducible representations (irrep) labeled by the $z$-component angular momentum $j=\pm \frac{1}{2},\pm \frac{3}{2},...,\pm (\left \lfloor{\frac{n+1}{2}}\right\rfloor - \frac{1}{2})$, where $\left \lfloor{x}\right \rfloor$ is the floor function. Note that for $n\in$ odd, the irreps with $j=\pm \frac{n}{2}$ are essentially the same. Since the Hamiltonian along the $k_z$ axis 
\bea
h(0,0,k_z)=\bigoplus_{j} h_j(k_z)
\eea
is block-diagonal in irrep $j$, we can define an independent topological charge $Q_j$ for each block Hamiltonian $h_j(k_z)$. \\

These topological charges $\{\tilde{Q}_j\}$ are defined as follows. For a given irrep $j$, the Hamiltonian $h_j(k_z)$ at each $k_z$ can be viewed as a 0d system with $\tilde{C}_n$ symmetry. 
Since both $h_j(0)$ and $h_j(\pi)$ are gapped (as the Dirac nodes are away from TRIMs), we can define a quantity
\bea
\omega_j(k_i) = \frac{1}{2}[N_{j}^c(k_i)-N_{j}^v(k_i)]
\eea
for each of them, which measures the number difference between the filled states $N_{j}^v(k_i)$ and the empty states $N_{j}^{c}(k_i)$ at $k_i=0,\pi$ for irrep $j$.
When $\omega_j(0)\neq\omega_j(\pi)$, there necessarily exist gapless modes between 0 and $\pi$ along $k_z$ that cannot be removed without closing the gaps at $k_i$. The number of the gapless modes is simply given by the topological charge\cite{kobayashi2015topological,yang2015topological} 
\bea
\tilde{Q}_j = \omega_j(0)-\omega_j(\pi). 
\label{Eq: Topological charge}
\eea
When $\tilde{Q}_j=0$, $h_j(0)$ and $h_j(\pi)$ are topologically equivalent and one can generally find an adiabatic path along $k_z$ to smoothly deform $h_j(0)$ to $h_j(\pi)$ without closing the energy gap, and hence with no gapless points in between. 
When $\tilde{Q}_j\neq 0$, there necessarily exist $|\tilde{Q}_j|$ of gapless 1d unidirectional modes along the $k_z$ axis that are either left movers ($\tilde{Q}_j<0$) or right movers ($\tilde{Q}_j>0$). 

To determine the number of robust Dirac points from the topological charges $\{\tilde{Q}_j\}$, we need to consider two additional contraints. 
First, in the presence of inversion and time-reversal symmetries, the spectra of $h_{j}(k_z)$ and $h_{-j}(k_z)$ are degenerate. The two irreps $\pm j$ therefore have the same topological charge  
\bea
\tilde{Q}_j = \tilde{Q}_{-j}.
\label{Eq: Topological Charge_2d irrep}
\eea
Second, the definition of a Dirac point requires the number of left movers and that of right movers to be the same, which imposes a ``charge conservation" condition 
\bea
\sum_j \tilde{Q}_j = 0.
\label{Eq: topological charge conservation}
\eea

With the constraints in Eq. \ref{Eq: Topological Charge_2d irrep} and Eq. \ref{Eq: topological charge conservation}, the number of indepedent topological charges for a $C_n$-invariant system is (i) $n/2-1$ when $n$ is even ; (ii) $(n-1)/2$ when $n$ is odd. Our choice of independent topological charges is given by:
\bea
\{\tilde{Q}_{\frac{1}{2}},\tilde{Q}_{\frac{3}{2}},...,\tilde{Q}_{\frac{n-1}{2}-1}\},\ \ &&\text{for }n\in\text{even} \nonumber \\
\{\tilde{Q}_{\frac{1}{2}},\tilde{Q}_{\frac{3}{2}},...,\tilde{Q}_{\frac{n}{2}-1}\},\ \ &&\text{for }n\in\text{odd}.
\label{Eq: Indepedent topological charges}
\eea
Since the Dirac points always come in pairs, the total number of bulk Dirac points (DP) along $k_z$ axis is given by
\bea
\text{Number of DPs} = 2\sum_{j} |\tilde{Q}_j|,
\eea
where only indepedent $\tilde{Q}_j$s defined in Eq. \ref{Eq: Indepedent topological charges} will be counted. We emphasize that a nontrivial $\tilde{Q}_j$ is a necessary but insufficient condition for the existence of bulk Dirac points. For example, when a pair of Dirac points is formed by the band inversion of the same set of bands and both live in $k_z\in[0,\pi]$, $\tilde{Q}_j$ is incapable of diagnosing them.

\subsubsection{Topological Charge for Dirac Superconductors}
\label{Subsubsec: Topo charge for DSC}

We now generalize the concept of topological charges to describe rotational symmetric DSCs in the weak pairing limit.
Consider a superconductor whose normal state is invariant under an $n$-fold rotational operation $\tilde{C}_n$, and the pairing gap transforming under $\tilde{C}_n$ as 
\bea
\tilde{C}_n \Delta({\bf k}) \tilde{C}^T = e^{i\frac{2\pi}{n}\alpha} \Delta(C_n{\bf k}), 
\label{Eq: gap Rotation}
\eea
where $\alpha = 0,1,2,...,n-1$.
We can define an $n$-fold rotational operation in the Nambu basis for the corresponding BdG Hamiltonian
\bea
C_n = \begin{pmatrix}
	\tilde{C}_n & 0 \\
	0 & e^{i\frac{2\pi}{n}  \alpha }\tilde{C}_n^*, 
\end{pmatrix}
\label{Eq: BdG Rotation}
\eea
such that the BdG Hamiltonian is invariant under $C_n$. 

Due to the presence of both time-reversal and inversion symmetries, every electron band belonging to irrep $j$ is grouped with another electron band with $-j$ to form a 2d irrep, whose particle-hole partners are thus a pair of degenerate hole bands with $\pm(\alpha-j)$. Therefore, a Dirac-like gapless crossing among such two pairs of degenerate BdG bands can only be achieved if $\alpha\neq 0$, when the electron and hole bands belong to different irreps.
Moreover, as we previously discussed in Sec. \ref{Subsec: Symmetry for BdG Systems}, the compatibility relation with TRS requires 
\bea
e^{i\frac{2\pi}{n}\alpha} = -1,
\eea
which implies that $\alpha = \frac{n}{2} \in \mathbb{Z}$. This leads to the following two choices of $n$ and $\alpha$ to achieve a time-reversal DSC: (i) $n=4,\ \alpha=2$; (ii) $n=6,\ \alpha=3$. In the weak pairing limit, the topological charge for such a DSC can be defined by summing over the $\tilde{Q}_j^{(e)}$ from the electrons and the $\tilde{Q}_j^{(h)}$ from the holes. In particular,

\begin{itemize}
	\item $n=4$, $\alpha=2$: For the electron part, there are four inequivalent irreps $j=\pm\frac{1}{2},\pm\frac{3}{2}$, based on which a single indepedent topological charge $\tilde{Q}_{\frac{1}{2}}^{(e)}$ is well-defined. Since PHS will transform the electron state $|\pm\frac{1}{2},e\rangle$ and $|\pm \frac{3}{2},e\rangle$ to the hole state $|\pm \frac{3}{2},h\rangle$ and $|\pm\frac{1}{2},h\rangle$, respectively, the topological charge for the holes can be related to that of the electrons by 
	\bea
	\tilde{Q}_{\frac{1}{2}}^{(h)} = - \tilde{Q}_{\frac{3}{2}} ^{(e)} = \tilde{Q}_{\frac{1}{2}} ^{(e)}.
	\eea
	The total topological charge for the entire BdG system is given by
	\bea
	Q_{\frac{1}{2}} = \tilde{Q}_{\frac{1}{2}}^{(e)} + \tilde{Q}_{\frac{1}{2}}^{(h)} = 2 Q_{\frac{1}{2}}^{(e)}.
	\label{Eq: C4 topological charge}
	\eea
	When $Q_{\frac{1}{2}}\neq 0$, the BdG system is a DSC protected by $C_4$ symmetry.
	
	\item $n=6$, $\alpha=3$: For $\tilde{C}_6$ symmetry, there are six irreps $j=\pm\frac{1}{2},\pm\frac{3}{2},\pm\frac{5}{2}$. As a result, one can define two independent topological charges $\tilde{Q}_{\frac{1}{2}}^{(e)}$ and $\tilde{Q}_{\frac{3}{2}}^{(e)}$ to characterize the Dirac nodes for the electron part. While PHS transforms the electron states $|\pm\frac{1}{2},e\rangle$ and $|\pm \frac{5}{2},e\rangle$ to the hole states $|\pm \frac{5}{2},h\rangle$ and $|\pm\frac{1}{2},h\rangle$, respectively, the electron states $|\pm \frac{3}{2},e\rangle$ are transformed into $|\pm\frac{3}{2},h\rangle$. Therefore, we have
	\bea
	\tilde{Q}_{\frac{1}{2}}^{(h)} =  \tilde{Q}_{\frac{1}{2}} ^{(e)} + \tilde{Q}_{\frac{3}{2}} ^{(e)},\ \ \tilde{Q}_{ \frac{3}{2}}^{(h)} = - \tilde{Q}_{\frac{3}{2}} ^{(e)}.
	\eea
	As a result, the topological charges for such a BdG system are given by
	\bea
	Q_{\frac{1}{2}} &=& \tilde{Q}_{\frac{1}{2}}^{(e)} + \tilde{Q}_{\frac{1}{2}}^{(h)} = 2\tilde{Q}_{\frac{1}{2}}^{(e)} + \tilde{Q}_{\frac{3}{2}}^{(e)} \nonumber \\
	Q_{\frac{3}{2}} &=& \tilde{Q}_{\frac{3}{2}}^{(e)} + \tilde{Q}_{\frac{3}{2}}^{(h)} = 0.
	\label{Eq: C6 topological charge}
	\eea
	For $Q_{\frac{1}{2}}\neq 0$, since the BdG Dirac point is formed between electron bands with $j=\pm \frac{1}{2}$ and hole bands with $j=\pm\frac{5}{2}$, it is actually a ``double" Dirac point with a linear dispersion along $k_z$ and quadratic in-plane dispersions \cite{ffnote3}. By definition, such a BdG system is dubbed a {\it double DSC}.
\end{itemize}

\subsubsection{From Dirac Semimetal to Dirac Superconductor}
\label{Subsubsection: DSM to DSC}

The relations in Eq. \ref{Eq: C4 topological charge} and Eq. \ref{Eq: C6 topological charge} reveal deep connections between DSMs and DSCs. In particular, starting from a DSM with a non-trivial $\tilde{Q}_j$, the final SC state is guaranteed a DSC if $Q_{\frac{1}{2}} \neq 0$.

As an example, let us consider a $\tilde{C}_6$-symmetric DSM with a pair of bulk Dirac points, which consists of $|j=\pm\frac{1}{2}\rangle$ and $|j=\pm\frac{3}{2}\rangle$. Without loss of generality, we assume the topological charges to be $Q_{\frac{1}{2}}^{(e)} =-Q_{\frac{3}{2}}^{(e)} =1$. By developing time-reversal-symmetric superconductivity with $\alpha=3$, we follow Eq. \ref{Eq: C6 topological charge} and arrive at
\bea
Q_{\frac{1}{2}} =1,\  Q_{\frac{3}{2}} =0.
\eea
Therefore, the final SC state is a double DSC with a pair of double Dirac points along $k_z$ axis. 

Similarly, we can apply this topology analysis to study the possibility of DSC phase for a superconducting DSM with other rotation symmetries or irreps. A summary of these results is listed in Table. \ref{Table: From DSM to DSC}.

\subsection{Inversion Symmetry Indicator $\kappa_{2d}$}
\label{Subsec: inversion symmetry indicator}

In this subsection, we discuss how to diagnose the existence of inversion-protected 1d Majorana hinge modes in a 3d DSC phase. 
This is in fact not a straight-foward task since DSC is a gapless phase, and the classification and an indicator that can diagnose boundary modes for such a 3d gapless superconductor with time-reversal and inversion symmetries are still unknown to the best of our knowledge. Given that the Dirac nodes are generally away from the high-symmetry plane $k_z=0$, here we propose to diagnose the existence of Majorana hinge modes in a 3d DSC by studying the inversion-protected topology in its $k_z=0$ plane, which is effectively a 2d gapped class-DIII system with a 2d inversion symmetry \cite{C3}.

We do so by using a $\mathbb{Z}_4$ inversion symmetry indicator $\kappa_{2d}$ for 2d gapped inversion-symmetric DIII superconductors. 
A 2d indicator of such kind was first conjectured in Ref. \cite{hsu2019inversion} for a half-filled case, generalized later in several works \cite{skurativska2019atomic,ono2018symmetry}, and recently one of us and collaborator have shown that the classification is indeed $\mathbb{Z}_4$ and that $\kappa_{2d}$ used in this work can indeed faithfully diagnose the boundary modes  based on a combined $K$ group and real-space topological crystal analysis \cite{AHSS}.
Such a boundary diagnostic depends only on the inversion eigenvalues of occupied BdG bands at the four high-symmetry points, and can detect whether a given 2d superconductor has no boundary modes, Majorana edge modes, or Majorana corner modes. 

The inversion operator for a BdG Hamiltonian is defined as follows.  
For a superconductor whose superconducing gap transforms under the normal-state inversion $\tilde{\cal P}_k$ as $\tilde{\cal P}_k\Delta(k)\tilde{\cal P}_k^T = \eta \Delta(-k)$, we can define an inversion operator in the Nambu basis 
\bea
{\cal P}_k =
\begin{pmatrix}
	\tilde{\cal P}_k & 0 \\
	0 & \eta \tilde{\cal P}_k
\end{pmatrix}
\eea
such that the BdG Hamiltonian is invariant under $\cal{P}$$_k$. 
Here $\eta=\pm 1$ indicates that the superconducting gap is parity-even or odd. 

With the BdG inversion operator ${\cal P}$ defined, we can now write down the inversion symmetry indicator 
\begin{align}
\kappa_{2d}=\frac{1}{2}\sum_k(N^+[H(k)]-N^+[H^0(k)])~~~~{\text{mod}}~4,  
\end{align}
where $N^+[h]$ denotes the number of even-parity occupied bands in a given Hamiltonian $h$, and the factor $1/2$ accounts for the Kramers degeneracy. Here, $H(k)$ is the BdG Hamiltonian of interest with $2\tilde{N}$ bands at a high-symmetry point $k=\Gamma,X,Y,M$, and $H^0(k)$ is a reference BdG Hamiltonian with the same number of bands $2\tilde{N}$. This latter term with $H^0(k)$ removes the contribution to the former term that orignates from the $k$-dependent phase factors carried by the inversion operator $\cal{P}$$_k$ itself. In the rest of this paper, we will take $H^0(k)=[I_{\tilde{N}},-I_{\tilde{N}}]$ regardless of the actual form of the given BdG Hamiltonian. 
Specifically, while $\kappa_{2d}=$1 and 3 correspond to first-order topological phases with Majorana edge modes \footnote{The $\kappa_{2d}=3$ phase exhibits coexisting Majorana edge and corner modes, but the corner modes are expected to be buried in the edge modes and are hard to detect.}, $\kappa_{2d}=2$ corresponds to a higher-order 2d strong phase featuring two inversion-protected Majorana Kramers pairs, one on each of the opposite corners.


We now turn back to the 3d nodel superconductors. For a given 3d DSC $H_{dsc}({\bf{k}})$, since its $k_z=0$ plane is gapped, we can compute $\kappa_{2d}$ for $H_{dsc}(kz=0)$. When $\kappa_{2d}=2$, we expect corner Majorana Kramers pairs on the $k_z=0$ plane, and consequently two-fold degenerate Majorana hinge modes in the spectrum of $H_{dsc}({\bf{k}})$. These hinge modes originate from extending the $k_z=0$ corner Majoranas in the $k_z$ direction up to the $k_z$ planes where the 2d bulk gap closes, i.e. the planes where the Dirac nodes lie within. 

In the weak-pairing limit, we can further relate this BdG indicator $\kappa_{2d}$ to an analogous inversion symmetry indicator $\tilde{\kappa}_{2d}$ defined for a normal state as \cite{ono2018symmetry}
\begin{align}
\kappa_{2d} = (1-\eta) \tilde{\kappa}_{2d}, 
\label{sc2normal}
\end{align}
where $\eta=1$ and $-1$ are for even- and odd-parity superconductors, respectively. 
However, a 2d normal state with both time-reversal and inversion symmtries has a $\mathbb{Z}_2$ classification, just like a 2d class-AII system without inversion. This means that the inversion symmetry does not induce new phases additional to the ones protected by time-reversal symmetry. This $\mathbb{Z}_2$ indicator is therefore just the familiar $\mathbb{Z}_2$ topological index $\nu$ for a time-reversal-invariant normal state. 

For even-parity superconductors ($\eta=1$), it is clear that $\kappa_{2d}$ is always zero because of the exact cancellation between the contribution to $\tilde{\kappa}_{2d}$ (or $\nu$) from the electrons and from the holes. We therefore should restrict ourselves to odd-parity superconductors ($\eta=-1$) to search for topologically non-trivial phases. 
In particular, to achieve a HOTDSC, a good starting point according to Eq. \ref{sc2normal} is a topological normal state whose $k_z=0$ plane has $\nu=1$. As we further introduce odd-parity pairing, we will obtain a $\kappa_{2d}=2$ superconducting state at $k_z=0$ with inversion-protected corner Majoranas, which further indicates the existence of 1d hinge Majorana modes in the 3d DSC.


\subsection{Mirror Chern Number ${\cal C}_M$}
\label{Subsec: mirror Chern number}
In this subsection, we discuss the mirror symmetry $M_z$ that sends $z\rightarrow -z$ and the corresponding mirror Chern number ${\cal C}_M$ for BdG systems. 
For our purpose of achieving a robust DSC from a $\tilde{C}_n$-symmetric DSM, we focus on the cases where $n=4,6$ with $\alpha=n/2$, as concluded in section \ref{Subsubsection: DSM to DSC}.
Importantly, since such a DSC is always invariant under a two-fold rotational $C_2$ and spatial inversion ${\cal P}$, it automatically has the mirror symmetry $M_z=C_2\cal{P}$. 
It is therefore important to study the corresponding mirror Chern number ${\cal C}_M$, since $M_z$ can protect {\it unwanted} gapless surface states (for our purpose) when ${\cal C}_M$ is non-zero, as we show in the following. 

For a $C_n$-symmetric BdG Hamiltonian, $M_z$ can be defined in the Nambu basis as 
\bea
M_z = (C_n)^{\frac{n}{2}} {\cal P} &=& \begin{pmatrix}
	\tilde{C}_2 & 0 \\
	0 & (-1)^{\alpha}\tilde{C}_2^*
\end{pmatrix} 
\begin{pmatrix}
	\tilde{\cal P} & 0 \\
	0 & \eta \tilde{\cal P}
\end{pmatrix} \nonumber \\
&=& \begin{pmatrix}
	\tilde{M}_z & 0 \\
	0 & (-1)^{\alpha} \eta \tilde{M}_z^*
\end{pmatrix} \nonumber \\
&=& \begin{pmatrix}
	\tilde{M}_z & 0 \\
	0 & (-1)^{\alpha} \tilde{M}_z
\end{pmatrix},
\label{Eq: BdG Mirror}
\eea
where $\tilde{C}_2$, $\tilde{\cal{P}}, \tilde{M}_z$ are the two-fold rotation, inversion, and mirror operators for normal state Hamiltonian repectively. In the last step of Eq. \ref{Eq: BdG Mirror}, we make use of the fact that $\tilde{M}_z^* = -\tilde{M}_z$ for spinful fermions and $\eta=-1$ for odd-parity pairing.

Given a normal state  with an electronic mirror Chern number ${\tilde{\cal C}}_M$, the ${\cal C}_M$ for the BdG system is 
\bea
{\cal C}_M = [1+ (-1)^{\alpha}] \tilde{\cal C}_M = \left\{
\begin{array}{lr}
	2 \tilde{\cal C}_M \ \ \ \alpha \in \text{even}, \\
	0 \ \ \ \ \ \ \ \alpha \in \text{odd}. \\
\end{array}
\right.
\eea
This relation implies that $M_z$-protected 2d surface states can only exist when $\alpha$ is even. To achieve a HOTDSC, where the 2d surface has to be gapped, we are therefore limited to the case with $n=6$ and $\alpha=3$. 

\subsection{A Recipe for HOTDSC}
\label{Subsec: topological requirements for HOTDSC}

\begin{table*}[t]
	\includegraphics[width=\linewidth]{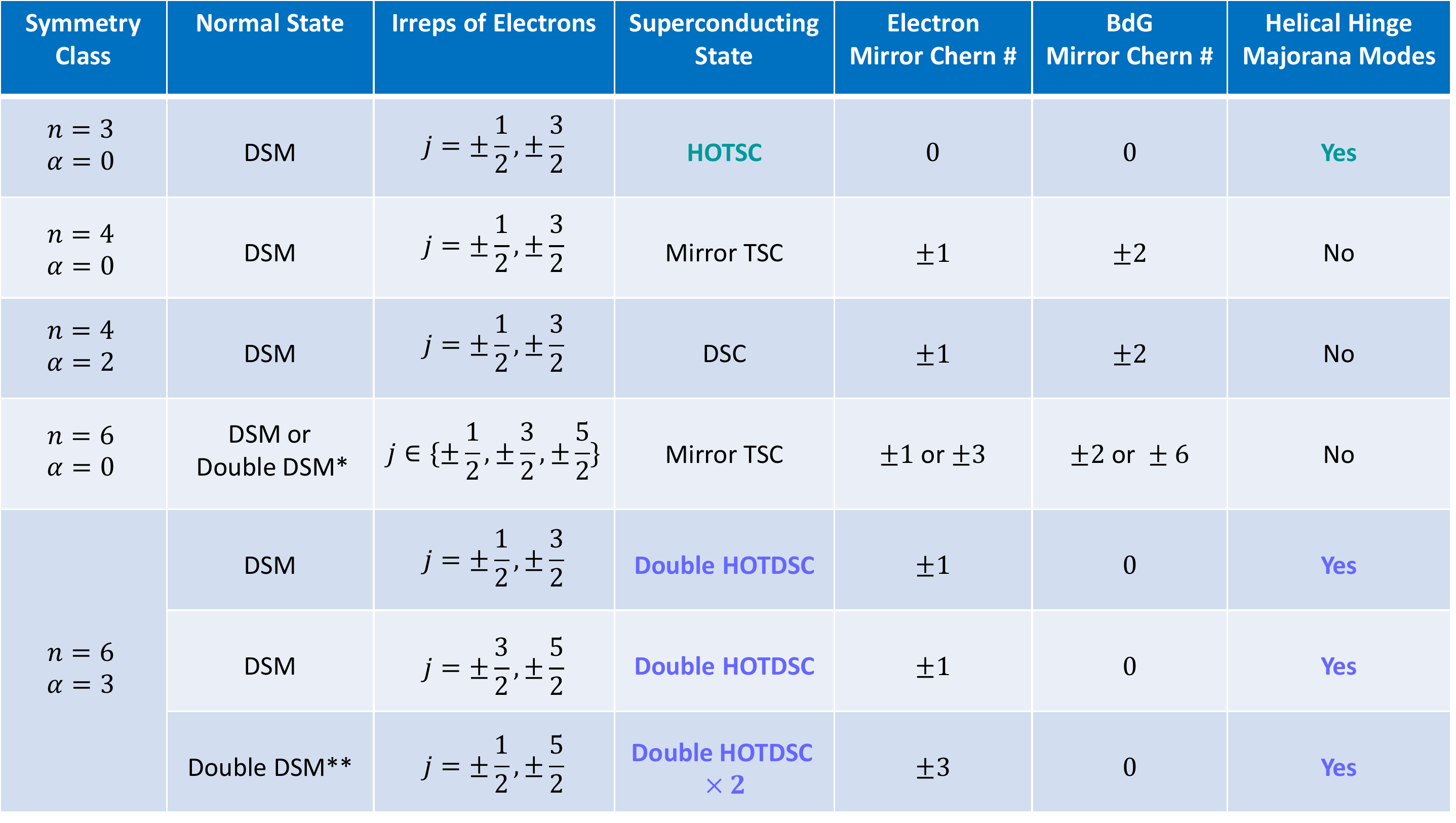}
	\caption{Summary on topological properties for superconducting DSMs with an $n$-fold rotation symmetry. For our purpose, we have only considered odd-parity pairing terms for DSMs that preserve TRS. The pairing term is even (odd) under $n$-fold rotation symmetry when $\alpha=0$ ($\alpha=n/2$). The presence of hinge Majorana modes indicates that the BdG system is higher-order topological. \\
		$*$ For $n=6,\alpha=0$, the normal state is a DSM if $j=\pm\frac{1}{2},\pm\frac{3}{2}$ or $j=\pm\frac{3}{2},\pm\frac{5}{2}$, and the superconducting state is a mirror TSC with $|{\cal C}_M|=2$. For $j=\pm\frac{1}{2},\pm\frac{5}{2}$, the normal state is a double DSM and the superconducting state has $|{\cal C}_M|=6$ \\
		$**$ The resulting superconducting state hosts two pairs of double Dirac nodes in the BdG spectrum.}
	\label{Table: From DSM to DSC}
\end{table*}

As we discussed above, the band topology of a superconducting Dirac semimetal is characterized by the following topological indices:
\bea
\{Q_j, {\cal C}_M, \kappa_{2d} \},
\eea 
which characterize the existence of 3d rotation-protected bulk Dirac nodes, 2d mirror-protected Majorana surface states, and 1d inversion-protected hinge Majorana modes, respectively. Based on our topological index analysis, we have summarized the possible superconducting states for DSMs with time-reversal-invariant odd-parity pairing in Table \ref{Table: From DSM to DSC}. 

In particular, starting from a $\tilde{C}_3$-symmetric DSM, odd-parity pairing can spoil the bulk Dirac physics and drive the system into a higher-order TSC with inversion protected hinge Majorana modes. With $\tilde{C}_4$ symmetry, the superconducting DSM could be (i) a mirror-protected TSC with gapped bulk and Majorana surface states when $\alpha=0$; (ii) a DSC with bulk Dirac nodes and Majorana surface states when $\alpha=2$. Similar mirror TSC phase can be achieved for a $\tilde{C}_6$-invariant DSM when the pairing satisfies $\alpha=0$. Finally, one of our main findings is that a HOTDSC phase is {\it only} possible in a $\tilde{C}_6$-symmetric superconducting DSM whose pairing is odd under both inversion and six-fold rotation.

\section{Model}
\label{Sec: Model}

In this section, we construct a minimal $\textbf{k}\cdot\textbf{p}$ model and the corresponding tight-binding model on a 3d hexagonal lattice that realize the HOTDSC phase predicted by our general topological index analysis in the previous section. Our models for a HOTDSC phase exhibit the expected set of bulk topological indices and the defining properties in the spectrum, namely the gapless bulk Dirac nodes, gapped surface states, and gapless Majorana hinge modes, as we demonstrate analytically and numerically below in the continuum and lattice models respectively. As we show in the following, our numerical and analytical results are completely consistent with each other, providing compelling support for our predictions on HOTDSC.

\subsection{Tight-binding Model for $C_6$-symmetric DSM}
\label{Subsec: lattice minimal model}

\subsubsection{The minimal ${\bf k\cdot p}$ model for DSM}

We start with a four-band minimal continuum model for a 3d DSM that was considered in Ref.  \cite{wang2012dirac} and \cite{wang2013three},
\bea
h_0 = v(k_x \gamma_1 - k_y\gamma_2) + m(k) \gamma_5
\label{Eq: DSM k dot p}
\eea
up to ${\cal O}(k^2)$ order, where the mass term is given by $m(k) = M_0 -M_1 (k_x^2 + k_y^2) - M_2 k_z^2$. 
Here, the Dirac matrices are defined as 
\bea
\gamma_1 &=& s_z\otimes \sigma_x, 
\gamma_2 = s_0\otimes \sigma_y, \nonumber \\
\gamma_3 &=& s_x\otimes \sigma_x, 
\gamma_4 = s_y\otimes \sigma_x, \nonumber \\
\gamma_5 &=& s_0\otimes \sigma_z, 
\eea
where $s_i$ and $\sigma_i$ denote the Pauli matrices in the spin  $s=\uparrow,\downarrow$ and orbital $\sigma=s,p$ bases, respectively.
The four basis states can therefore be labeled by their $z$-component angular momenta as $|\frac{1}{2}\rangle,|\frac{3}{2}\rangle,|-\frac{1}{2}\rangle,|-\frac{3}{2}\rangle$. 
In such spin and orbital bases, the spatial inversion symmetry and time reversal symmetry are given by
\bea
\tilde{\cal P} &=& s_0 \otimes \sigma_z = \gamma_5 \nonumber \\
\tilde{\Theta} &=& is_y \otimes \sigma_0 K = i\gamma_{13} K, 
\eea
where $\gamma_{ij}\equiv [\gamma_i,\gamma_j]/(2i)$. 
The continuum model respects a continous rotational symmetry around the $z$-axis 
\bea
\tilde{C}_{\infty} = e^{i\theta J_z}
\eea
for an arbitrary rotation angle $\theta$, where the generator is a diagonal matrix $J_z = \text{diag}\{\frac{1}{2},\frac{3}{2},-\frac{1}{2},-\frac{3}{2}\}$. This rotational symmetry prevents avoided-crossings between bands of different irreps, and is thus the symmetry that protects gapless Dirac points. In particular, when $M_0M_2>0$, this model has two $\tilde{C}_{\infty}$-protected four-fold degenerate bulk Dirac points at 
\bea
k_x=k_y=0,\ k_z= \pm k_0 = \pm \sqrt{\frac{M_0}{M_2}}.
\eea
These bulk Dirac points remain robust even when we regularize the model on a lattice and break the continuous $\tilde{C}_{\infty}$ down to a discrete rotational symmetry $\tilde{C}_{n}$ with $n=3,4,6$, as we previously discussed in Sec. \ref{Subsubsec: DSM topo charge}.     

We also include the symmetry-allowed next leading order term $h_1({\bf k})$ for this DSM model in our later discussions \cite{wang2012dirac}. In particular,
\bea
h_1(\textbf{k}) = v_z \begin{pmatrix}
	0 & 0 & 0 & k_z k_-^2 \\
	0 & 0 & k_z k_-^2 & 0 \\
	0 & k_z k_+^2 & 0 & 0 \\
	k_z k_+^2 & 0 & 0 & 0 \\
\end{pmatrix},
\label{Eq: Fermi Arc Deformation}
\eea
which clearly vanishes along $k_z$ axis. Physically, $h_1(\textbf{k})$ has no effect on the bulk Dirac nodes but is able to deform the dispersion of Fermi arc surface states \cite{kargarian2016are}.

\subsubsection{The lattice model}

\begin{figure}[t]
	\centering
	\includegraphics[width=0.5\textwidth]{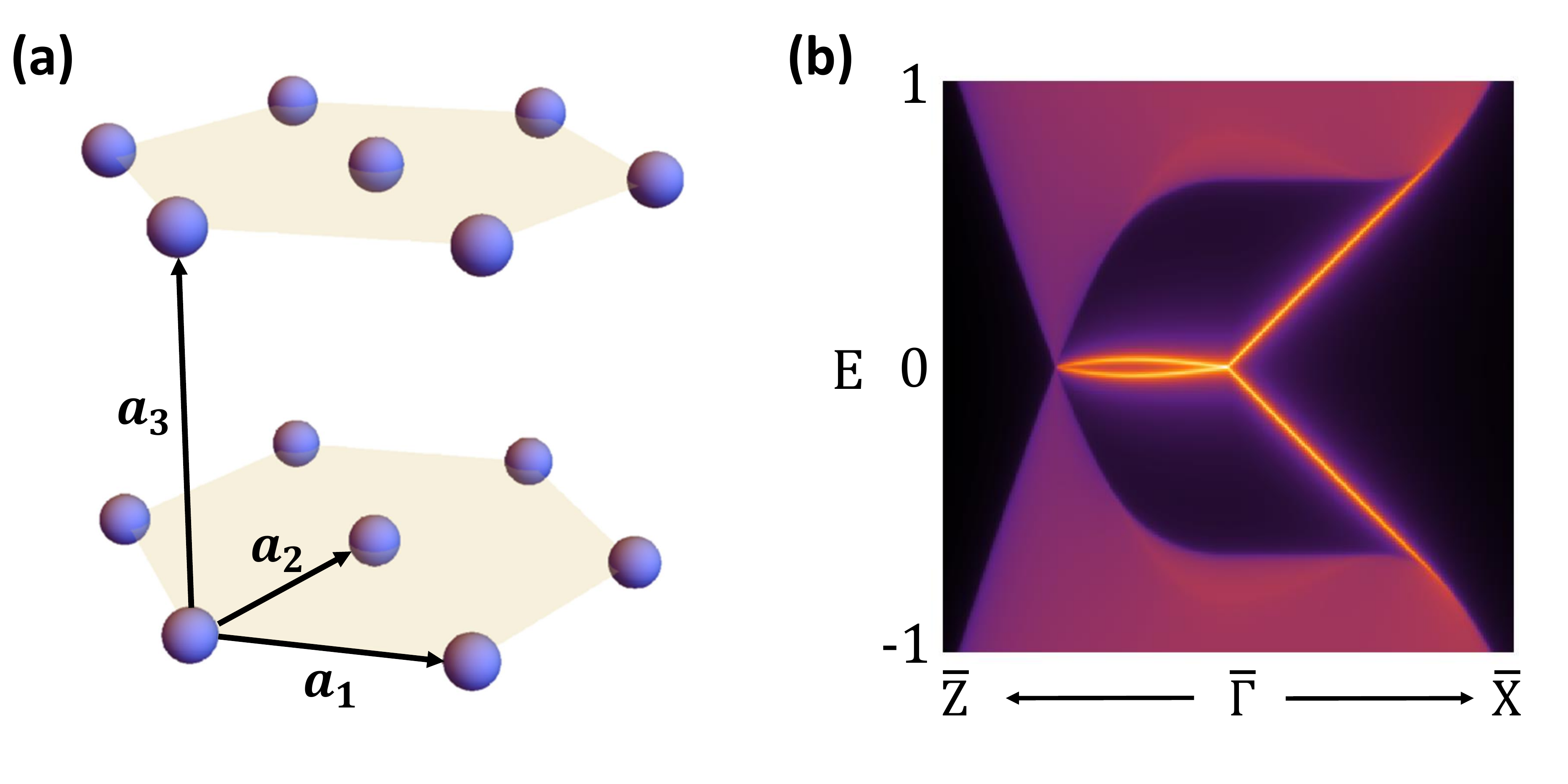}
	\caption{The hexagonal lattice structure and lattice vectors for the tight-binding model are shown in (a). The (010) surface spectrum of the lattice DSM model is plotted in (b), which clearly shows the bulk Dirac point and the Fermi-arc surface states. Here we choose $v_0 = 0.25, u_z=0.1, t=1, t_z =2, k_0 =0.3\pi$. The high-symmetry momenta in the surface Brillouin zone are defined as $\bar{\Gamma}=(0,0), \bar{X}=(\pi,0), \bar{Z}=(0,\pi)$, which follows the notation $(k_x, k_y)$.}
	\label{Fig: Lattice}
\end{figure}

We now regularize the continuum model and put it on a hexagonal lattice  characterized by the lattice vectors ${\bf a}_1 = (1,0,0)$, ${\bf a}_2 = \frac{1}{2}(1,\sqrt{3},0)$, and ${\bf a}_3 = (0,0,1)$,
as shown in Fig. \ref{Fig: Lattice} (a). The resulting tight-binding model has the form 
\bea
h_{tb} ({\bf k}) = \sum_{i=1}^{5} d_i({\bf k}) \gamma_i
\eea
with
\bea
d_1 &=& v_0 [2\sin k_1 +\sin (k_1-k_2) +\sin k_2] \nonumber \\
d_2 &=& \sqrt{3}v_0 [\sin k_2 - \sin (k_1-k_2)] \nonumber \\
d_3 &=& u_z\sin k_z [2\cos k_1 - \cos (k_1 - k_2) -\cos k_2 ]  \nonumber \\
d_4 &=& - u_z \sqrt{3} \sin k_z [\cos (k_1-k_2) -\cos k_2] \nonumber \\
d_5 &=& t [\cos k_1 + \cos k_2 + \cos (k_1-k_2)-3] \nonumber \\
&& + t_z (\cos k_z -\cos k_0),
\label{Eq: DSM tight binding}
\eea
where we have defined $k_1=k_x$ and $k_2=(k_x+\sqrt{3}k_y)/2$. In the long-wavelength limit, $h_{tb}$ reproduces the continuum model in Eq. \ref{Eq: DSM k dot p} and Eq. \ref{Eq: Fermi Arc Deformation} if we replace $v_0 \rightarrow \frac{4}{3} v$, $u_z \rightarrow \frac{4}{3}v_z$, $t \rightarrow \frac{4}{3} M_1$, $t_z \rightarrow 2M_2$, and  $k_0 \rightarrow \cos^{-1} (1- \frac{M_0}{2M_2})$.
Nonetheless, instead of the continuous rotational symmetry $\tilde{C}_{\infty}$, $h_{tb}$ preserves the six-fold rotational symmetry 
\bea
\tilde{C}_6 = e^{i\frac{\pi}{3}J_z}
\eea
such that $\tilde{C}_6 h_{tb}(k_1,k_2,k_z) \tilde{C}_6^{\dagger} = h_{tb} (k_1-k_2,k_1,k_z)$.
Together with the fact that there exists a pair of bulk Dirac points at ${\bf k}=(0,0,\pm k_0)$, we have shown that $h_{tb}$ is a lattice model that realizes a $C_6$-symmetric DSM. 

We numerically demonstrate the existence of bulk Dirac points along $\Gamma$-$Z$ and the corresponding Fermi arc states by plotting the (010) surface spectrum in a semi-infinite geometry using the iterative Green function method [see Fig. \ref{Fig: Lattice} (b)]. 
It is easy to check that the bulk Dirac points are labeled by the topological charge $Q_{\frac{1}{2}}^{(e)}=-Q_{\frac{3}{2}}^{(e)}=1$ (as discussed in Sec. \ref{Subsubsec: DSM topo charge}), and the Fermi arc surface states are protected by the mirror Chern number $\tilde{{\cal C}}_M=1$.

\subsection{Odd-Parity Superconductivity and Higher-Order Topology}
\label{Subsec: Odd parity pairing}

\begin{figure}[t]
	\centering
	\includegraphics[width=0.5\textwidth]{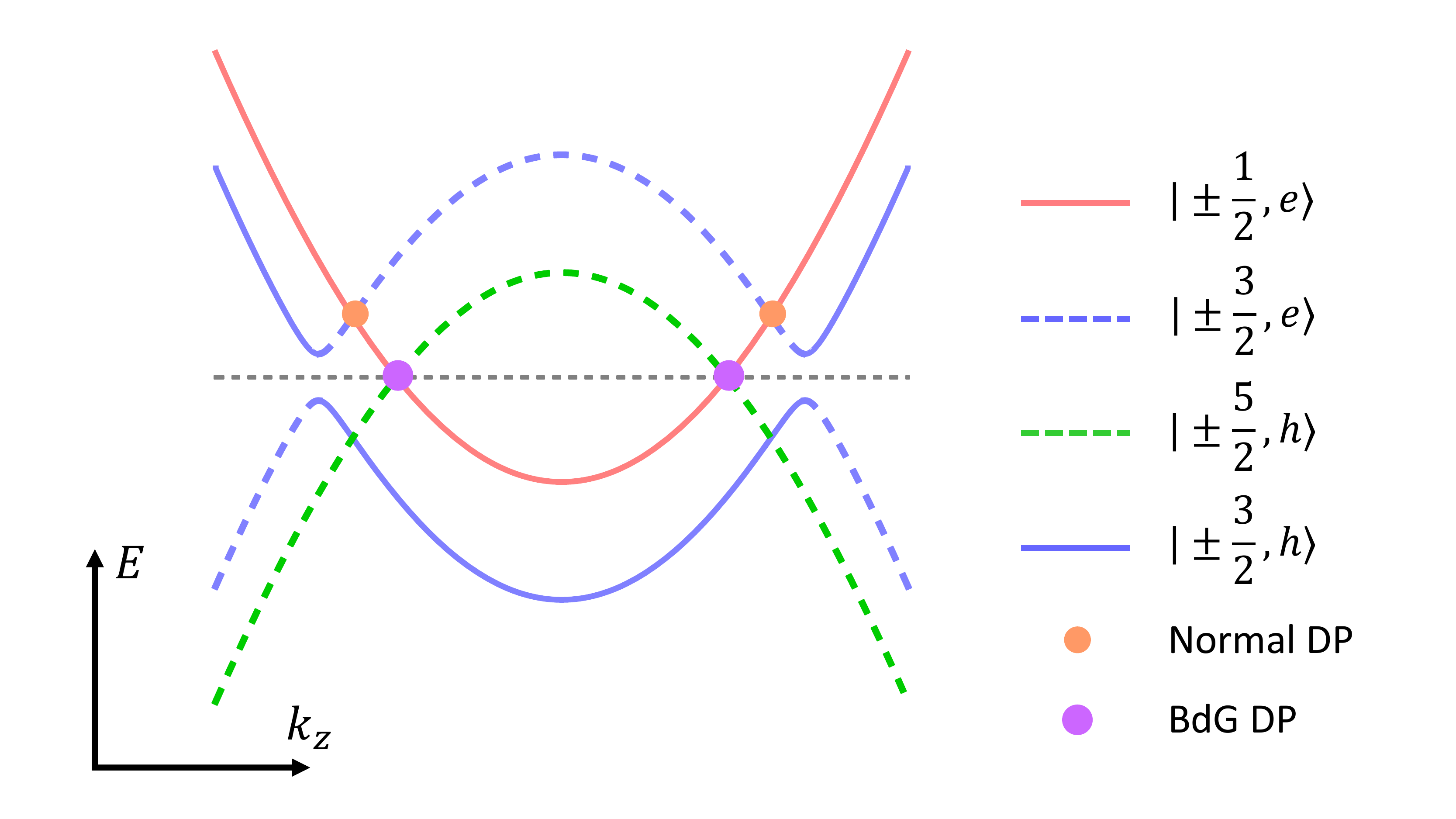}
	\caption{We schematically plot the BdG band structure along $k_z$ for our doped DSM model. The BdG (normal) Dirac nodes are shown in the purple (orange) dots. Notice the anti-crossing between $|\pm \frac{3}{2},e\rangle$ state (blue solid line) and $|\pm \frac{3}{2},h\rangle$ state (blue dashed line), which originates from the $p_z$-wave pairing in $H_{\Delta}({\bf k})$.}
	\label{Fig: Bulk Dirac Schematic}
\end{figure}

We now consider a doped $\tilde{C}_6$-symmetric DSM model and introduce symmetry-allowed pairing gaps to construct the BdG Hamiltonian $H(\textbf{k})=H_0(\textbf{k})+H_{\Delta}(\textbf{k})$ for a HOTDSC phase.

Starting from the continuum limit, the minimal Hamiltoniain for a doped DSM can be written in the Nambu basis as 
\bea
\label{H0conti}
H_0(\textbf{k}) &= v(k_x \tau_0 \otimes \gamma_1 - k_y\tau_z \otimes \gamma_2)\nonumber\\
&+ m(k) \tau_z \otimes \gamma_5-\mu\tau_z\otimes \gamma_0,
\eea
where $\mu$ denotes the chemical potential capturing the doping effect, and $\tau_i$ denote the Pauli matrices in the particle-hole basis. For our purpose, we will ignore the ${\cal O}(k^3)$ term in Eq. \ref{Eq: Fermi Arc Deformation} in $H({\bf k})$ for now and will include it as a perturbation in later discussions. 

To obtain a HOTDSC phase, the pairing term in $H_{\Delta}$, as already discussed in Sec. \ref{Sec: Symmetry indicators}, has to be time-reversal symmetric, but odd under both inversion and six-fold rotation. This requires the full Hamiltonian $H(k)$ to be invariant under the following symmetry operations defined in the Nambu basis:  
\bea
\Theta = i\tau_0\otimes \gamma_{13} K,\ \ 
{\cal I} = \tau_z\otimes \gamma_5,\ \ 
C_6 = \begin{pmatrix}
	\tilde{C}_6 & 0 \\
	0 & -\tilde{C}_6^* 
\end{pmatrix}. \nonumber \\
\eea
Given the above symmetry constraints, we find that the general symmetry-preserving pairing term that respects fermionic statistics has the form
\begin{align}
	H_{\Delta}(\textbf{k})&=\Delta_1 [k_z \tau_x\otimes s_z \otimes(\sigma_0-\sigma_z)]\nonumber\\
	&+\Delta_2 [(k_x^2-k_y^2) \tau_y\otimes s_y\otimes \sigma_x- 2 k_x k_y \tau_x \otimes s_x \otimes \sigma_y]\nonumber\\
	&+\Delta_3 [k_z (k_x \tau_y\otimes s_0\otimes \sigma_y + k_y\tau_x\otimes s_z\otimes \sigma_y)], 
	\label{Deltaconti}
\end{align}
up to ${\cal O}(k^2)$ order, where parameters $\Delta_{1,2,3}$ are assumed to be real for simplicity. Here, the $\Delta_2$ and $\Delta_3$ terms correspond to two distinct nodal d-wave pairings, and the $\Delta_1$ term corresponds to a $p_z$-wave pairing, with all of these terms being time-reversal symmetric and \textit{odd} under inversion and six-fold rotation (i.e. $\alpha=3$). Throughout our work, we will set $\Delta_3=0$ for simplicity since it is irrelevant to the topological physics we study. Importantly, the $p_z$-wave pairing exists only between electrons and holes with $j=\pm\frac{3}{2}$ to respect the rotational symmetry, which therefore leaves us two pairs of rotation-protected Dirac nodes between the BdG bands comprising mostly $j=\pm 1/2$ electron and $j=\pm 5/2$ hole bands respectively [see Fig.  \ref{Fig: Bulk Dirac Schematic} for a schemtic demonstration].

Recall that the normal state $H_0({\bf k})$ is a DSM with the following topological indices
\begin{align}
Q_{\frac{1}{2}}^{(e)} = - Q_{\frac{3}{2}}^{(e)} =1,\ \tilde{\cal C}_M=1,\ \nu=1.
\end{align}
Following the results in Sec. \ref{Sec: Symmetry indicators}, it is then straightforward to show that our BdG system $H({\bf k})$ hosts the following BdG indices
\bea
Q_{\frac{1}{2}}=1,\ Q_{\frac{3}{2}}=0,\ {\cal C}_M=0,\ \kappa_{2d}=2,
\eea
which are exactly the indices that correspond to a HOTDSC with a pair of bulk double Dirac nodes, gapped surface states, and hinge Majorana modes. As we have discussed the Dirac physics in the bulk, in the following, we will demonstrate that the hinges are gapless while the surface states are gapped by analytically constructing an effective surface theory (Sec. \ref{Subsec: Effective boundary theory}) and numerically calculating the boundary spectrum on an open geometry (Sec. \ref{Subsec: hinge Majorana}). 
	
\subsection{Effective Boundary Theory and Bulk-boundary Correspondence}
\label{Subsec: Effective boundary theory}

\begin{figure}[t]
	\centering
	\includegraphics[width=0.5\textwidth]{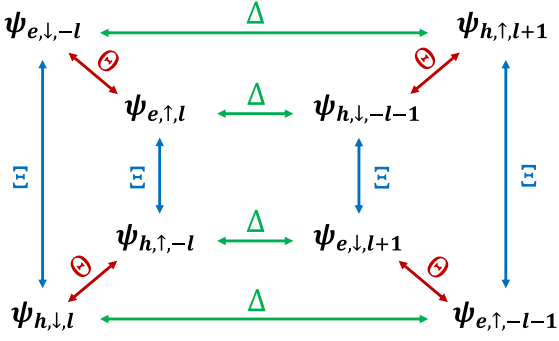}
	\caption{The relations among various low-energy surface fermion fields in the cylindrical geometry. The blue (red) arrows connect the fermion fields that are particle-hole (time-reversal) partners. The green arrows connect pairs of fermion fields that form Cooper pairs.}
	\label{Fig: Relation}
\end{figure}
	
Building an effective surface theory is a crucial step in understanding the higher-order bulk-boundary correspondence for HOTDSCs. For our purpose, we focus on the surface spectrum for an infinite long cylinder geometry. Here we work with the cylindrical coordinate $(r,\theta,z)$, where an open surface around the cylinder is labeled by the polar angle $\theta$. Because of the nontrivial bulk topological indices of our model, we expect topological features in the boundary spectrum. For example, if the surface state for our system has a $\theta$-dependent energy gap that vanishes at some special angles, this clearly suggests gapless hinge Majorana physics in our system. 

Our starting point is the continuum Hamiltonian $H_0({\bf k})$ on a cylindrical geometry, whose surface states will be solved analytically. In the weak pairing limit, we treat $h_1(\textbf{k})$ in the normal state and the pairing term $H_{\Delta}({\bf k})$ as perturbations, and we project them onto the low-energy bases spanned by the surface states of $H_0({\bf k})$ later. In the absence of the perturbations, the normal state Hamiltonian $H_0(\textbf{k})=\rm{diag}$$[h_{\uparrow}({\bf k}),h_{\downarrow}({\bf k}),-h_{\downarrow}({\bf k}),-h_{\uparrow}({\bf k})]$ is block-diagonal, where 
\bea
\label{Eq: Effective DSM model}
h_{\uparrow}({\bf k}) &=& \begin{pmatrix}
	m(k) & vk_+ \\
	vk_- & -m(k) \\
\end{pmatrix}.
\eea
Since $h_{\downarrow}(m(k),v)=h_{\uparrow}(m(k),-v)^*$, we only need to solve for the surface Fermi arc states for the $2\times 2$ Hamiltonian $h_{\uparrow}({\bf k})$. To transform $h_{\uparrow}({\bf k})$ into the cylindrical coordinate, we write $k_+=e^{i\theta}(k_r+ik_{\theta})$ and $k_-=e^{-i\theta}(k_r-ik_{\theta})$, where $k_r=-i\frac{\partial}{\partial r}$ and $k_{\theta}=-i\frac{1}{r}\frac{\partial}{\partial \theta}$. 
For a large system whose radius $r$ is much larger than the lattice constant $a$, 
we can further make the approximation $k_x^2+k_y^2=k_r^2+k_{\theta}^2-i\frac{1}{r}k_r \approx k_r^2 + k_{\theta}^2$. 
To solve for the eigenstates localized on the 2d open surface at $r=R$, we write down an ansatz wavefunction 
\bea
\psi_l(r,k_z,\theta) = {\cal N} e^{ik_z z} e^{il\theta} f(r) \xi(\theta), 
\label{Eq: ansatz wavefunction}
\eea
where ${\cal N}$ is the normalization factor and $l\in \mathbb{Z}$. Here $f(r)$ is the radial part of $\psi_l$ and $\xi(\theta)$ is a two-component spinor. With this ansatz, we arrive at the following surface-localized eigenstates and their eigen-energies following Ref.  \cite{imura2011spin,zhang2018crystalline}:
\bea
\psi_{e,\uparrow,l} &=& e^{il\theta} \begin{pmatrix}
	ie^{i\theta} \\
	1 \\
\end{pmatrix},\ \ \ \ \ \ E_{e,\uparrow,l} = \frac{v}{R}(l+\frac{1}{2}), \nonumber \\
\psi_{e,\downarrow,l} &=& e^{il\theta} \begin{pmatrix}
	-ie^{-i\theta} \\
	1 \\
\end{pmatrix},\ \ E_{e,\downarrow,l} = -\frac{v}{R}(l-\frac{1}{2}), \nonumber \\
\psi_{h,\uparrow,l} &=& e^{il\theta} \begin{pmatrix}
	-ie^{-i\theta} \\
	1 \\
\end{pmatrix},\ \ E_{h,\uparrow,l} = \frac{v}{R}(l-\frac{1}{2}), \nonumber \\
\psi_{h,\downarrow,l} &=& e^{il\theta} \begin{pmatrix}
	ie^{i\theta} \\
	1 \\
\end{pmatrix},\ \ \ \ \ \ E_{h,\downarrow,l} = -\frac{v}{R}(l+\frac{1}{2}).
\label{Eq: Fermi arc information}
\eea
Here we have dropped the spatial part $f(r)$ and the plane-wave factor $e^{ik_z z}$ for simplicity. As discussed in Ref. \cite{imura2011spin,zhang2018crystalline}, the electronic surface states $\psi_{e,\uparrow,l}$ and $\psi_{e,\downarrow,l}$ exist only between the bulk Dirac points at $k_z=\pm k_0$ and thus manifest themselves as the Fermi arc states. Interestingly, Eq. \ref{Eq: Fermi arc information} predicts the Fermi arc states having a finite-size gap of $\frac{v}{R}$, which is numerically confirmed in Appendix A. Physically, such finite-size gap will generally appear in the energy spectrum when placing Dirac fermions on the surface of a cylinder, which happens in topological insulator nanowires \cite{rosenberg2010wormhole} and Weyl/Dirac semimetal nanowires \cite{imura2011spin,zhang2018crystalline}. The physical origin of this factor has been attributed to the spin Berry phase accumulated while the Dirac fermion (e.g. TI surface state or Fermi arc state of semimetals) circulates around the cylinder, which can be canceled by inserting magnetic flux to the cylinder. So in this sense, this factor is a combined effect of the spin-texture of the Dirac fermion \cite{rosenberg2010wormhole,imura2011spin,zhang2018crystalline} and the curvature of the cylindrical geometry \cite{wieder2019strong}.

The next step is to project the perturbation terms onto the low-energy surface states to obtain an effective surface theory. To fully incorporate the time-reversal symmetry $\Theta$, 
the particle-hole symmetry $\Xi$, 
and the Cooper pairing $\Delta$, the minimal set of basis states of the effective theory necessarily consists of eight fermionic fields [see Fig. \ref{Fig: Relation}]. 
Nonetheless, we can simplify the basis by dividing these eight fields into 
\bea
\Phi_{1,l} &=& (\psi_{e,\uparrow,l},\psi_{e,\downarrow,-l},\psi_{h,\uparrow,l+1},\psi_{h,\downarrow,-l-1})^T, \nonumber \\
\Phi_{2,l} &=& (\psi_{e,\uparrow,-l-1},\psi_{e,\downarrow,l+1},\psi_{h,\uparrow,-l},\psi_{h,\downarrow,l})^T. 
\eea

Since the two sets of states are related by $\Phi_{1,l} = \Phi_{2,-l-1}$ and are decoupled from each other, we only need to construct the low-energy theory spanned by $\Phi_{1,l}$. 
In the basis of $\Phi_{1,l}$, the symmetry operations are given by
\bea
{\cal P}= (-1)^{l+1} \mu_0 \otimes s_0,\  \Theta = i\mu_0\otimes s_y K,\ 
C_6 = e^{i \frac{\pi}{3} {\cal J}_z}, \nonumber \\
\eea
where the rotation generator is given by 
\begin{eqnarray}
{\cal J}_z = \text{diag}\{l+\frac{3}{2}, -(l+\frac{3}{2}), l+\frac{5}{2},-(l+\frac{5}{2}) \}.
\end{eqnarray}

\begin{figure}[t]
	\centering
	\includegraphics[width=0.5\textwidth]{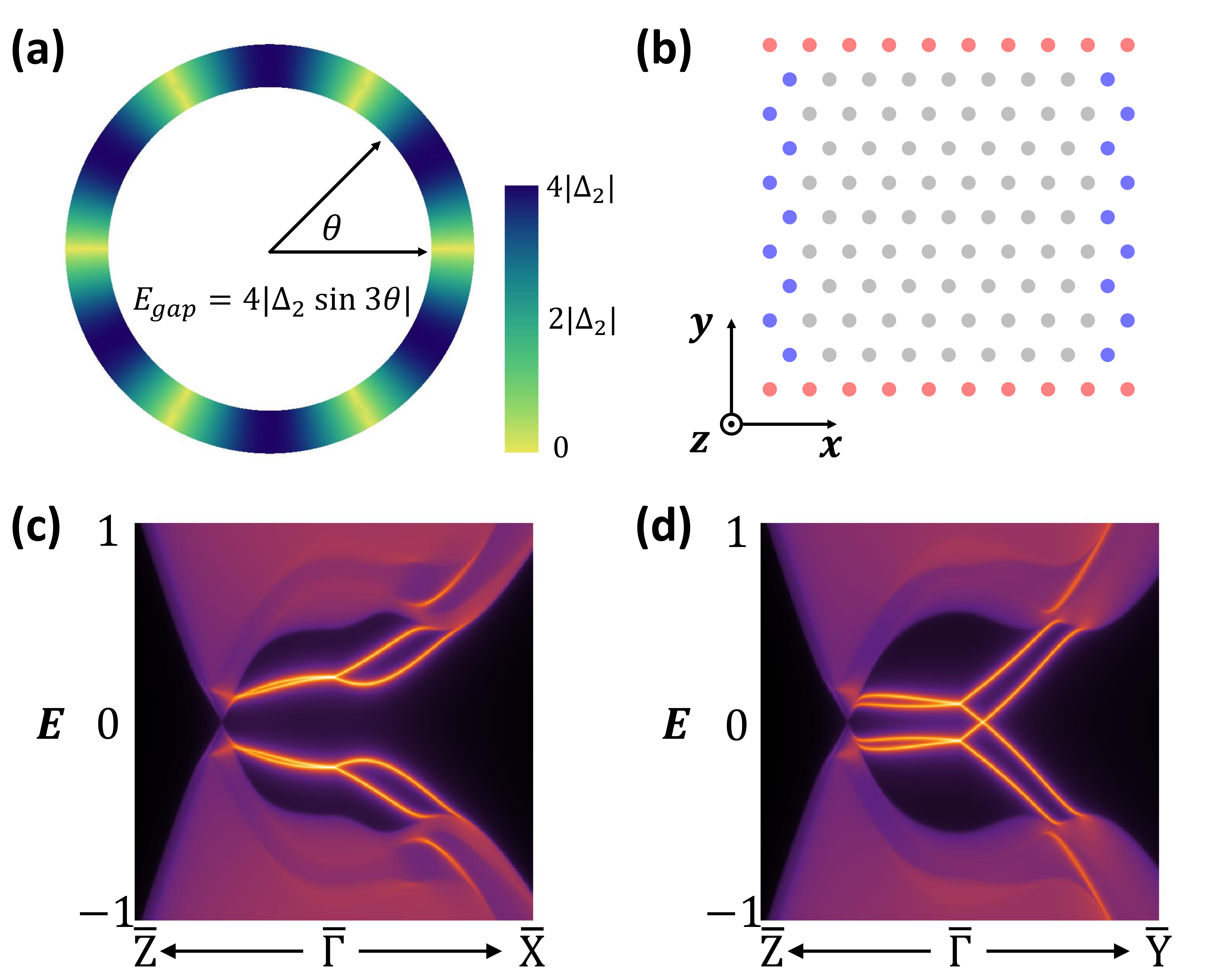}
	\caption{Surface physics of the HOTDSC phase. In (a), we schematically plot the effective pairing gap of the Fermi-arc surface states as a function of polar angle $\theta$, following the analytical result in Eq. \ref{Eq: Surface gap closing condition}. Numerically, as shown in (b), we consider two distinct surface terminations: (i) the smooth surface (red dots) with $\theta=\pi/2$ along $x$ direction; (ii) the rough surface (blue dots) with $\theta=0$ along $y$ direction. We calculate the surface state spectrum of HOTDSC for both the smooth and rough surfaces in (c) and (d), respectively. The gapless (gapped) surface state for the rough (smooth) surface agrees with our analytical prediction. }
	\label{Fig: Surface}
\end{figure}

By projecting $h_1(\textbf{k})$ in Eq. \ref{Eq: Fermi Arc Deformation} and pairing term $H_{\Delta}$ in Eq. \ref{Deltaconti} onto $\Phi_{1,l}$, the resulting effective Hamiltonian and its eigenvalues are given by 
\begin{widetext}
	\bea
	&H_{\text{eff},l} =& \frac{v}{R}(l+\frac{1}{2}) + \begin{pmatrix}
		0 & -iv_zk_z e^{-i(2l+3)\theta} & \Delta_1 k_z e^{i\theta} & \Delta_2 e^{-i(2l+1)\theta}\sin 3\theta \\
		iv_zk_z e^{i(2l+3)\theta} & 0 & -\Delta_2 e^{i(2l+1)\theta}\sin 3\theta & -\Delta_1 k_z e^{-i\theta} \\
		\Delta_1 k_z e^{-i\theta} & -\Delta_2 e^{-i(2l+1)\theta}\sin 3\theta & 0 & iv_zk_z e^{-i(2l-1)\theta} \\
		\Delta_2 e^{i(2l+1)\theta}\sin 3\theta & -\Delta_1 k_z e^{i\theta} & -iv_zk_z e^{i(2l-1)\theta} & 0 \\
	\end{pmatrix} \label{Eq: Boundary Hamiltonian} \\
	&E_l (k_z, \theta) =&  \frac{v}{R}(l+\frac{1}{2}) \pm \sqrt{(v_zk_z\sin 3\theta)^2 + (\sqrt{(2\Delta_2\sin 3\theta)^2 + (v_z k_z \cos3\theta)^2} \pm \Delta_1k_z)^2}, 
	\label{Eq: Boundary Eigen-energy}
	\eea
\end{widetext}
where we have taken the large $R$ limit with $k_{\pm} \approx e^{\pm i\theta} k_r$, since $k_{\theta}$ is of ${\cal O}(\frac{1}{R})$. The dispersion relation $E_l (k_z, \theta)$ generally shows a finite energy gap for $k_z\neq 0$. On the other hand, the energy gap at $k_z=0$ is given by
\bea
E_\text{gap} (\theta) = 4 |\Delta_2 \sin 3\theta|.
\label{Eq: Surface gap closing condition}
\eea
In Fig. \ref{Fig: Surface} (a), we schematically show the spatial profile of the surface gap function $E_\text{gap}$ as a function of $\theta$ around the cylinder. It is clear that the surface gap vanishes only at {\bf six} special angles with
\bea
\theta = \frac{n \pi}{3},\ n=0,1,2,3,4,5, 
\eea
which directly suggests the existence of zero-energy hinge Majorana modes that live in-between the projected bulk Dirac nodes along $k_z$, if the system is placed in an open geometry.

\subsection{Numerical Evidence for HOTDSC phase}
\label{Subsec: hinge Majorana}
	
		\begin{figure*}[t]
		\centering
		\includegraphics[width=0.99\textwidth]{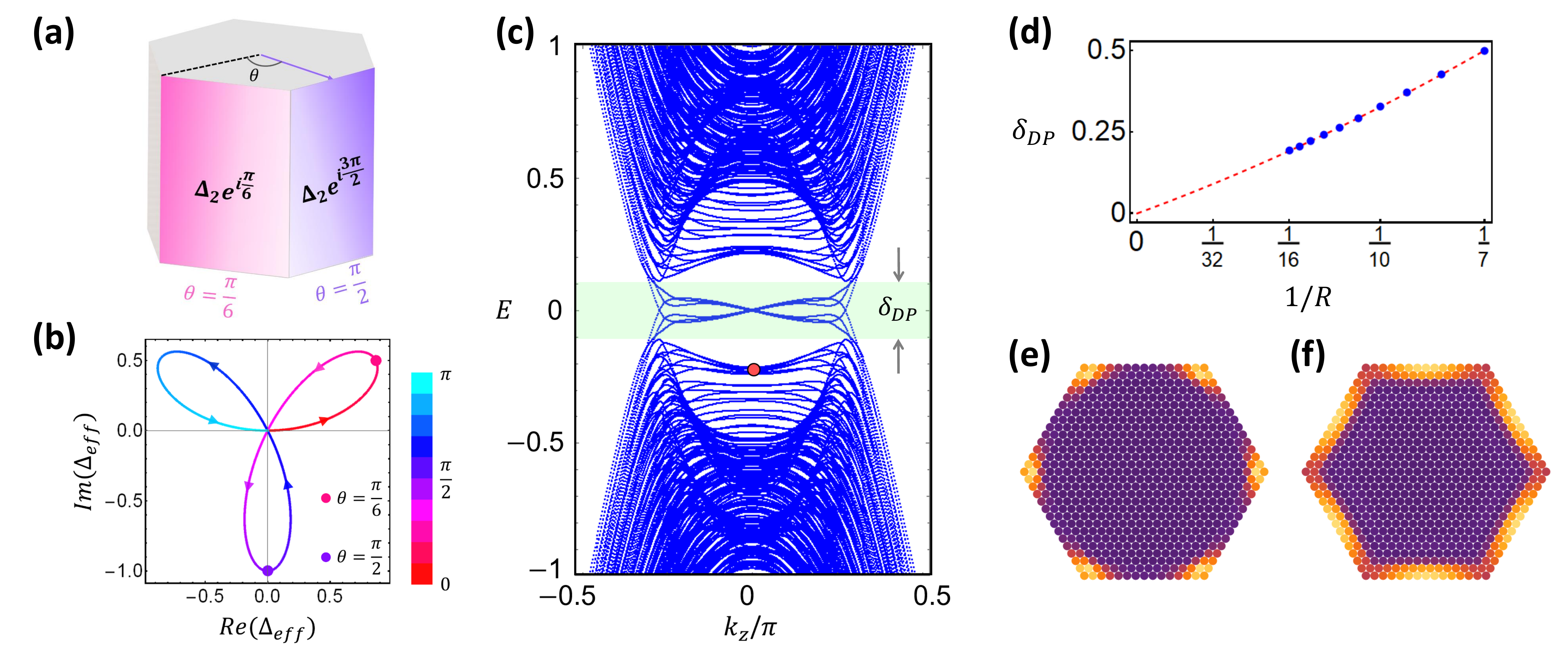}
		\caption{In (a), we schematically plot the hexagonal prism geometry of the HOTDSC model and label the pairing function $\Delta_s$ on two neighboring surfaces with $\theta=\pi/6$ and $\theta=\pi/2$. In (b), we plot the evolution of effective surface pairing $\Delta_s$ as a function of $\theta$ in the 2d plane spanned by the real and imaginary parts of $\Delta_s$. It is important to notice that the trajectory necessarily crosses the origin while evolving from $\theta=\pi/6$ (red dot) to $\theta=\pi/2$ (purple dot), which is guaranteed by the higher-order topology. In (c), we calculate the energy dispersion of our HOTDSC model in the infinitely long hexagonal prism geometry, which clearly reveals the gapless bulk Dirac nodes, a surface pairing gap near $k_z=0$, and the in-gap hinge Majorana modes between the bulk nodes. The scaling behavior of the finite size gap $\delta_{DP}$ of the bulk Dirac nodes is shown in (d). By calculating the spatial profile of wavefunctions in the prism cross section, we verify (i) the hinge nature of the zero-energy Majorana modes at $k_z=0$ in (e); (ii) the surface nature of the low-energy gapped state at $k_z=0$ (the red dot) in (f).}
		\label{Fig: Hinge Mode}
	\end{figure*}

Next, we numerically verify the HOTDSC physics in our model. The first step is to verify the analytical prediction for surface physics from the previous subsection. We will then proceed to directly show the coexistence of bulk Dirac nodes and hinge Majorana modes by numerically calculating the energy spectrum in a hexagonal prism geometry, which will unambiguously demonstrate the higher-order Dirac nature of our model. 

We start by regularizing our continuum BdG model (including $H_{\Delta}$) onto a 3d hexagonal lattice and arrive at
\bea
H_\text{TB}({\bf k}) &=& d_1 \tau_0 \otimes \gamma_1 + d_2\tau_z\otimes \gamma_2 + d_3 \tau_0 \otimes \gamma_3 \nonumber \\
&& + d_4\tau_z\otimes \gamma_4 + d_5 \tau_z \otimes \gamma_5 - \mu \tau_z \otimes I_4 \nonumber \\
&& + d_6\tau_y\otimes \gamma_4 + d_7 \tau_x \otimes \gamma_{35} \nonumber \\
&& + d_8 \tau_x \otimes (\gamma_5 - \gamma_{12}),
\eea
where we have defined
\bea
d_6 ({\bf k}) &=& \Delta_2 [2\cos k_1 - \cos (k_1 - k_2) -\cos k_2 ] \nonumber \\
d_7 ({\bf k}) &=& -\sqrt{3} \Delta_2 [\cos (k_1-k_2) -\cos k_2] \nonumber \\
d_8 ({\bf k}) &=& \Delta_1 \sin k_z.
\eea

Next, we numerically verify the surface gap closing condition in Eq. \ref{Eq: Surface gap closing condition} by solving for the boundary-localized eigenstates of our minimal lattice model on a 3d hexagonal lattice. Specifically, we consider the two semi-infinite configurations for side surfaces shown in Fig. \ref{Fig: Surface} (b): (i) the rough surface (colored in blue) along $y$ direction with $\theta=0$; (ii) the smooth surface (colored in red) along $x$ direction with $\theta=\pi/2$. We then calculate the surface-state spectra for both the smooth and the rough surfaces with $\Delta_1=0.1$ and $\Delta_2=0.25$, as shown in Fig. \ref{Fig: Surface} (c) and (d) respectively. We find that both surface calculations show one single bulk Dirac point at zero energy in the presence of finite pairing parameters, as indicated by the topological indices of our model. In particular, the smooth surface acquires a finite surface gap while the rough surface ($\theta=0$) remains gapless even in the presence of finite superconducting pairing \cite{fote1}, which validates our analytical prediction for the projected pairing gap on the surfaces in Eq. \ref{Eq: Surface gap closing condition}.

To directly reveal the hinge Majorana modes, it is necessary to place our model in an infinite long hexagonal prism geometry along $z$ direction, as shown in Fig. \ref{Fig: Hinge Mode} (a). For our purpose, all six side surfaces around the infinite prism are taken to be ``smooth surfaces", which are defined by the polar angle 
\bea
\theta_s=\frac{(2n+1)\pi}{6},\ \ n=0,1,...,5.
\eea
According to Eq. \ref{Eq: Surface gap closing condition}, all six surfaces  share a surface pairing gap that is proportional to $|\Delta_2|$, which just follows our calculation in Fig. \ref{Fig: Surface} (c). As an example, we focus on two adjacent surfaces: the pink surface with $\theta=\pi/6$ and the purple surface with $\theta=\pi/2$, which are shown in Fig. \ref{Fig: Hinge Mode} (a). The projected pairing function that controls the surface gap is simply the off-diagonal term in Eq. \ref{Eq: Boundary Hamiltonian}:
\bea
\Delta_s(\theta) = \Delta_2 e^{-i(2l+1)\theta}\sin 3\theta.
\eea
Interestingly, the surface pairing $\Delta_s$s for the pink and purple surfaces in Fig. \ref{Fig: Hinge Mode} (a) differ by a phase of $4\pi/3$ for the $l=0$ surface state. Therefore, rather surprisingly, the neighboring smooth surfaces are {\bf not} forming a surface mass domain wall with a $\pi$-phase difference. Although the $\pi$-phase domain wall physics serves as the key to understanding the boundary physics in many higher-order topological systems, the predicted hinge Majorana modes in our HOTDSC system seems to arise from a different origin.

To resolve the origin of hinge Majorana modes, we consider a 2d parameter space spanned by the real and imaginary parts of $\Delta_s(\theta)$ projected onto the surfaces. As we change the value of $\theta$ from $0$ to $\pi$, the possible value of $\Delta_s$ for the $l=0$ surface state is {\it constrained} to the 1d closed loop trajectory shown in Fig. \ref{Fig: Hinge Mode} (b). To demonstrate, we label the surface gaps $\Delta_s$ for the pink and purple surfaces in Fig. \ref{Fig: Hinge Mode} (a) by pink and purple dots in Fig. \ref{Fig: Hinge Mode} (b), respectively.  While in general gapped hinges are expected for a trajectory that avoids the origin of the parameter space, the trajectory in our case {\it necessarily} goes through the origin and thus enforces the existence of gapless hinge modes.

From a different perspective, we can adiabatically rotate the surfaces with polar angles $\theta=\frac{\pi}{6}$ and $\tilde{\theta}=\frac{\pi}{6}$ (in Fig. 6a) to $\theta'=\frac{\pi}{3}- \delta \theta$ and $\tilde{\theta}'=\frac{\pi}{3}+ \delta \theta$, respectively, which will not close the surface gaps on both surfaces. As $\delta \theta \rightarrow 0$, the surface mass for these two neighboring surfaces now becomes $\Delta(\theta') \sim -\delta\theta$ and $\Delta(\tilde{\theta}') \sim +\delta\theta$, which recovers a $\pi$-phase difference that was ``hidden" in the hexagonal prism geometry and thus explains the origin of the hinge modes.

Now we are ready to perform numerical calculation on the energy spectrum in the same infinite hexagonal prism geometry. The side length (or the ``radius") $R$ of the hexagonal cross section is taken to be 14 unit cells. In Fig. \ref{Fig: Hinge Mode} (c), we plot the energy spectrum of the hexagonal wire as a function of $k_z$ with the same set of parameters as that in Fig. \ref{Fig: Surface} (c). Just as we expect, at $k_z=0$, the surface Fermi arc state opens a finite pairing gap. We plot the in-plane spatial profile of the gapped state shown by the red dot in Fig. \ref{Fig: Hinge Mode} (f), which confirms its surface-state nature. Remarkably, inside the surface pairing gap, there exist six pairs of 1d Majorana channels. As shown in Fig. \ref{Fig: Hinge Mode} (e), we find one pair of hinge Majorana modes on each of the six hinges of the hexagonal prism. Notably, any attempt to eliminate these hinge Majorana modes will necessarily break the crystalline and TRS symmetries or close the bulk gap, which directly implies the intrinsic higher-order topology in the system.

Fig. \ref{Fig: Hinge Mode} (c) also reveals the bulk BdG Dirac nodes. Although the Dirac points appear to be gapped due to finite-size effects, the scaling behavior of the finite-size gap $\delta_{DP}$ in Fig. \ref{Fig: Hinge Mode} (d) clearly indicates the existence of gapless Dirac points in the thermodynamic limit. The red dashed line shows a polynomial-function fit of $\delta_{DP}$ as a function of the prism radius $R$ with
\bea
\delta_{DP} (R) = \frac{a}{R} + \frac{b}{R^2} + {\cal O}(\frac{1}{R^3})
\eea
and $a\approx b/2 \approx 2.7$. As expected, $\delta_{DP}(\infty)=0$ in the thermodynamic limit, confirming the gaplessness of the Dirac point in the system.

Therefore, the 3d bulk BdG Dirac physics and the coexisting hinge Majorana physics together establish our system as a higher-order topological Dirac superconductor protected by time-reversal symmetry, inversion symmetry, and six-fold rotational symmetry. The exotic boundary phenomena shown in the above numerical calculations originate from and agree with our theoretical symmetry indicator analysis. Hence, we have established the bulk-boundary correspondence of the HOTDSC phase in our minimal model.
	
\section{A Higher-Order Topological Quantum Critical Point}
	\label{Sec: QCP}
	
		\begin{figure}[t]
		\centering
		\includegraphics[width=0.49\textwidth]{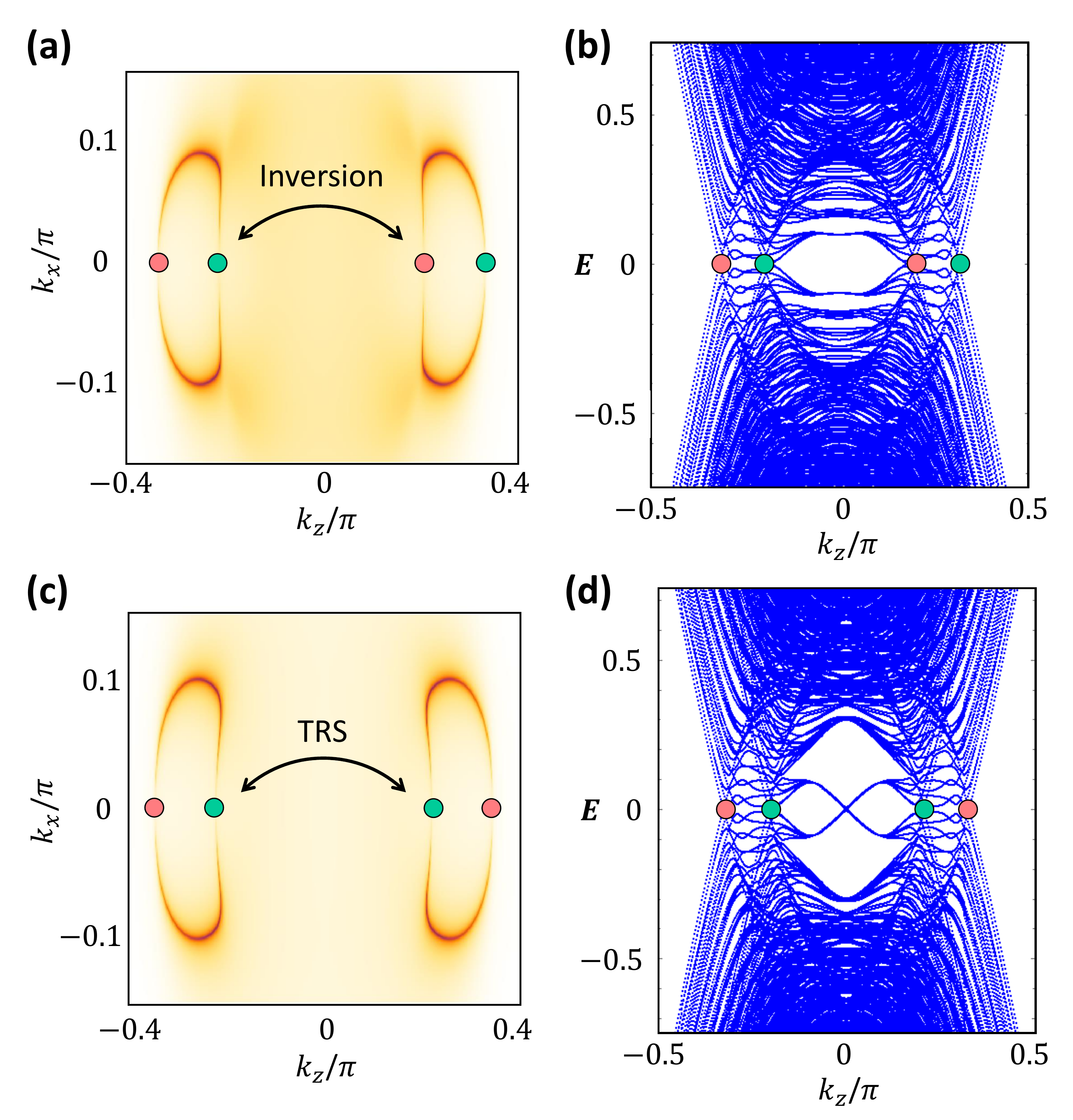}
		\caption{For the TRS breaking phase of a HOTDSC, we plot its Fermi surface at zero energy for the smooth (010) surface in (a), which clearly reveals the Majorana Fermi arcs connecting the bulk Weyl nodes (red and green dots). The color in these plots denotes the intensity of density of states at zero energy. This confirms the TRS breaking phase as a Weyl superconductor, which agrees with the energy spectrum calculation performed in the infinite prism geometry in (b). The Fermi surace of inversion breaking phase for (010) surface is plotted in (c). In addition to the bulk Weyl nodes and the Majorana Fermi arc states, the in-gap hinge Majorana modes show up around $k_z=0$ in the infintie prism geometry in (d), which establishes the inversion breaking phase as a higher-order Weyl superconductor.}
		\label{Fig: Higher order Weyl SC}
	\end{figure}
	
	In this section, we establish the HOTDSC phase as a higher-order topological quantum critical point, which could be driven into various (higher-order) topological phases through either spontaneous or explicit symmetry breaking, as shown in Fig. \ref{Fig: Schematic} (b). In particular, we define and construct a higher-order version of Weyl superconductors, a concept which has not before been discussed in the literature. This exotic higher-order gapless phase could be achieved by breaking inversion symmetry in a HOTDSC. In the following, we will discuss in details the resulting topological phases upon various symmetry breakings. The concepts of higher-order Weyl superconductor and higher-order topological quantum critical point arising from HOTDSC are among our important new theoretical findings.
	
	\subsection{$\Theta$ Breaking: Double Weyl Superconductor}
	When TRS is explicitly or spontaneously broken, each bulk Dirac node will immediately split into a pair of bulk Weyl nodes that carry monopole charges of $\pm 2$ in momentum space. Such Weyl nodes are often known as the {\it double Weyl points}, whose effective Hamiltonian generally takes the following form
		\begin{equation}
		h_\text{dW} = \begin{pmatrix}
		v_z k_z & v_{\parallel} k_+^2 \\
		v_{\parallel} k_-^2 & -v_z k_z \\
		\end{pmatrix}
		\end{equation}
	which hosts a quadratic in-plane dispersion, while keeping the dispersion linear along $k_z$.
	In addition, the hinge Majorana modes are expected to develop a Zeeman-like gap due to TRS breaking, which trivializes the higher-order topology of the system. Therefore, we expect that the broken TRS of a HOTDSC phase should lead to a {\it double Weyl superconductor}, which features double Weyl nodes in the bulk and Majorana Fermi-arc states on the surface.
	
	We now numerically demonstrate the emergence of double Weyl superconductivity upon TRS brekaing. To model the effect of TRS breaking, we add a generalized Zeeman-like TRS-breaking term to our minimal lattice model for HOTDSC:
	\bea
	H_z = g_0 \tau_z \otimes s_z \otimes \sigma_0 + g_1 \tau_z \otimes s_z \otimes \sigma_z.
	\eea
	In Fig. \ref{Fig: Higher order Weyl SC} (a) we show the Fermi surface for the spectrum of the (010) surface with $g_0=0.1,g_1=0.2$, where the red and green dots denote the bulk Weyl nodes with a monopole charge of $+2$ and $-2$, respectively. In addition, the Majorana Fermi arc states connecting the Weyl nodes always come in pairs, which is a signature of a Weyl superconductor with higher monopole charges. 
	
	We now examine the fate of hinge Majorana modes under TRS breaking by calculating the energy spectrum in an infinite hexagonal prism geometry [see Fig. \ref{Fig: Higher order Weyl SC} (b)]. In contrast to the spectrum without TRS breaking in Fig. \ref{Fig: Surface} (c), now the spectrum shows a finite energy gap at $k_z=0$, which implies the absence of gapless hinge modes. Moreover, upon the splitting of a Dirac node into a pair of Weyl nodes with opposite charges, a continuous distribution of gapless Majorana Fermi-arc surface states emerge between each pair of Weyl nodes, which is consistent with our finding in Fig. \ref{Fig: Higher order Weyl SC} (a). We therefore conclude that TRS breaking will indeed trivialize the higher-order topology and drive the system into a ``conventional" double Weyl superconductor without any gapless Majorana hinge modes.
	
\subsection{${\cal P}$ Breaking: Higher-Order Weyl Superconductor}
Breaking inversion symmetry ${\cal P}$ also drives a transition from a bulk Dirac node to a pair of Weyl nodes, but it is known that a weak ${\cal P}$ breaking term does not necessarily gap out the hinge Majorana modes. Although the hinge Majorana modes are no longer directly protected by the inversion symmetry, we expect that they can still exist as a result of extrinsic higher-order topology as long as the surfaces remain gapped around $k_z=0$. 

To confirm this picture, we add the following inversion-breaking perturbation
\bea
H_\text{ivb} = w \sin k_z \tau_z \otimes s_z \otimes \sigma_z,
\eea
to our lattice model, which preserves both TRS and $C_6$ symmetry. The Fermi surface plot for the (010) surface is shown in Fig. \ref{Fig: Higher order Weyl SC} (c), where we set $w=0.4$. Both the bulk Weyl nodes and the Majorana Fermi arc states are clearly revealed, which confirms the Weyl superconductor nature of the inversion-breaking phase. It is worth pointing out that the Weyl points here are split from a Dirac point in a different way compared with that in the TRS breaking phase, as shown in Fig. \ref{Fig: Higher order Weyl SC} (a) and (c). This is because, given a Weyl point, inversion symmetry (TRS) will enforce the existence of another Weyl point at an opposite crystal momentum with an opposite (same) monopole charge. We show explicitly how symmetries constrain the charges of Weyl points in Fig. \ref{Fig: Higher order Weyl SC} (a) and (c).

We then numerically calculate the energy spectrum of $H+H_\text{ivb}$ with $w=0.4$ in an infinitely long prism geometry. As shown in Fig. \ref{Fig: Higher order Weyl SC} (d), there exist bulk Weyl points and inter-node Majorana Fermi arc surface states as expected. Remarkably, we also find six pairs of hinge Majorana modes that live inside the pairing gap on the surface, similar to the case of HOTDSC. The coexisting 3d bulk Weyl points, 2d surface Fermi arc states, and 1d hinge Majorana states together define the ${\cal P}$-breaking phase as the first example of a truly exotic {\it higher-order Weyl superconductor}.

The topological features of this higher-order Weyl SC, including the bulk Weyl nodes, surface states and the hinge modes, are all separated in momentum space. It is therefore, in principle, possible to detect them individually in momentum-resolved spectroscopic experiments. Since the appearance of Majorana Fermi arc states is guaranteed by the bulk Weyl nature of the system, a higher-order Weyl SC by definition represents an example of hybrid higher-order topology \cite{bultinck2019three}, where 2d and 1d boundary modes coexist.

\subsection{$C_6$ Breaking: Higher-Order TSC}

When the $C_6$ symmetry is broken, the bulk Dirac points are expected to be gapped due to the absence of any symmetry protection. Nonetheless, the remaining symmetries, i.e., time-reversal and space-inversion, will still guarantee the stability of the Majorana hinge modes. We therefore arrive at a fully gapped higher-order TSC protected by inversion symmetry, which is expected to be also characterized by an inversion symmetry indicator $\kappa_{2d}=2$. In fact, the hinge Majorana modes will still be robust even if the inversion symmetry is weakly broken. This will lead to a TRS protected extrinsic higher-order TSC, as demonstrated in Fig. \ref{Fig: Schematic} (b). \\

Given all these resulting higher-order phases upon different symmetry breaking patterns, we expect the higher-order topology in the HOTDSC phase to be immune to generic weak non-magnetic disorder in the system even if such disorder violates $C_6$ or inversion symmetries.
Practically speaking, even if a material candidate fails to fulfill all the required symmetries for the HOTDSC phase listed in Sec. \ref{Sec: Symmetry indicators}, this ``failed" HOTDSC candidate could still be one of the exotic topological phases discussed in this section.
	
\section{Conclusion}
\label{Sec: Conclusion}

In summary, we introduce a new gapless phase of matter featuring higher-order band topology, namely, the higher-order topological Dirac superconductor, whose defining properties are (1) symmetry-protected three-dimensional Dirac nodes, (2) absence of two-dimensional Fermi-arc states, and (3) symmetry-protected one-dimensional Majorana hinge modes. Such an exotic nodal paired state is therefore topologically distinct from traditional topological superconductors and previously proposed Dirac superconductors. 
We establish that such a phase can be realized under the protection of six-fold rotation, spatial inversion, and time-reversal symmetries in the presence of unconventional pairing that is odd under inversion and rotation. 
This HOTDSC phase can therefore be fully characterized by a corresponding set of topological indices defined for Bogoliubov-de Gennes Hamiltonians. 
Following the above symmetry criteria, we further construct a 3d minimal lattice model for a HOTDSC by introducing symmetry-allowed nodal pairings in a hexagonal Dirac semimetal.  
In particular, we verify that our model exhibits the expected topological invariants and numerically demonstrate the three defining properties of HOTDSC on an open geometry.

In terms of materials search for HOTDSC phases, there are two possible routes to pursue. One is to look for time-reversal symmetric nodal superconductors with centrosymmetric hexagonal lattices. The other is to look for Dirac semimetals with the same type of lattices that develop nodal superconductivity. 
We point out that heavy fermion compounds could be an appealing platform that offers promising candidates. 
For the first route, there are in fact a few Uranium-based superconductors with hexagonal space group symmetries that are known to have unconventional pairings \cite{Joynt2002The}. In particular, UNi$_2$Al$_3$ and UPd$_2$Al$_3$ have $C_6$ rotational symmetry (space group No. 191) \cite{Geibel1993Ground}, while UPt$_3$ has $C_6$ screw rotational symmetry (space group No. 194)  \cite{Joynt2002The}, which we expect to work as normal $C_6$ for our purpose. Moreover, while UPd$_2$Al$_3$ is often considered as a non-phonon \cite{Jourdan1999Superconductivity} nodal \cite{Caspary1993Unusual,Tou1995d,Matsuda1997Observation,Hiroi1997Thermal} superconductor, UNi$_2$Al$_3$ and UPt$_3$ are experimentally shown to be spin-triplet superconductors \cite{Ishida2002Spin,Joynt2002The} with possible point nodes \cite{Tou1997NMR,Joynt2002The,Gannon2015New,Sumita2018Unconventional}. 
As for the second route, despite that the rotational symmetry is four-fold and the existence of superconductivity is yet to be explored, it has been pointed out that heterostructures involving rare-earth Kondo insulators can lead to a Dirac semimetal phase \cite{Ok2017Magnetic}. We hope our theory will inspire more future efforts towards realizing higher-order topology in heavy fermion superconductors and beyond. We believe that HOTDSC should exist in nature (or could be synthesized in the laboratory) since all the individual ingredients for its existence have already been realized in different situations.
	
\section{Acknowledgment}
	R.-X.Z is indebted to Jiabin Yu and Fengcheng Wu for helpful discussions. This work is supported by Laboratory for Physical Sciences and Microsoft. R.-X.Z is supported by a JQI Postdoctoral Fellowship. 
	
	Upon finalizing our manuscript, we became aware of a related work \cite{wieder2019strong} that also discusses higher-order Dirac physics.\\

	\appendix

		\section{Surface Gap Scaling of cylindrical Dirac semimetal}	    
	
		In this appendix, we discuss the finite-size effect of the surface Fermi arc states when a DSM is placed in a cylindrical (hexagonal prism) geometry. In the cylindrical geometry, the energy spectrum of Fermi arc surface states in the DSM model follows Eq. \ref{Eq: Fermi arc information}, which shows a finite-size gap between $E_{e,\uparrow,0}$ and $E_{e,\uparrow,-1}$:
		\bea
		\delta_\text{analytical} = E_{e,\uparrow,0} - E_{e,\uparrow,-1} = \frac{v}{R}.
		\eea
		Physically, this finite energy gap originates from the spin Berry phase effect of the Fermi arc states \cite{imura2011spin}. 
		
\begin{figure}[H]
			\centering
			\includegraphics[width=0.3\textwidth]{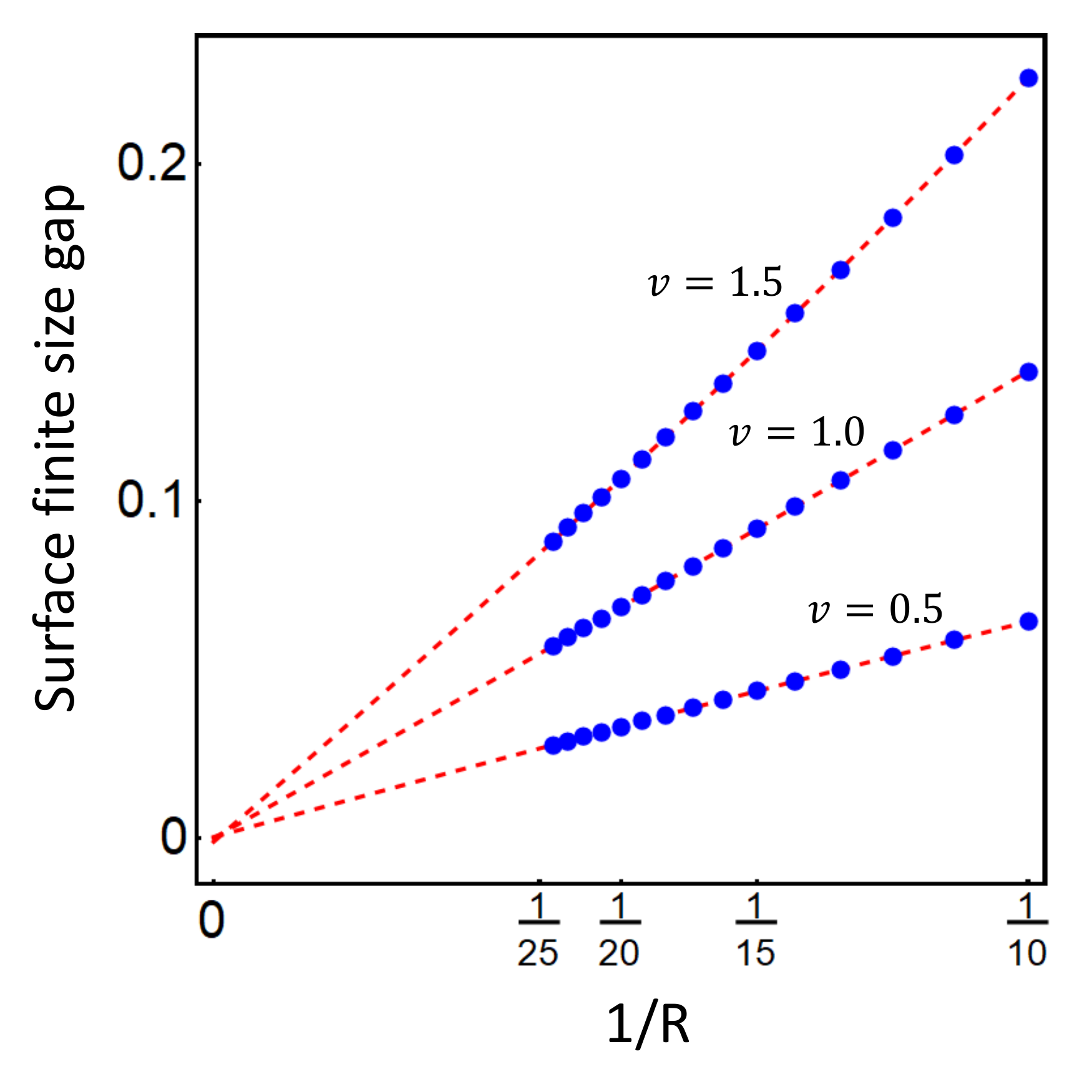}
			\caption{The surface gap scaling of a DSM with different $v$ is plotted in a hexagonal prism geometry by varying radius $R$.}
			\label{Fig: Finite size effect}
		\end{figure}

		As shown in Fig. \ref{Fig: Finite size effect}, we numerically calculate the surface gap of the DSM model at $k_z=0$ for $v\in\{0.5,1.0,1.5\}$ in a hexagonal prism geometry by changing the radius $R$. By performing polynomial fittings, we arrive at a simple universal relation between the numerical finite-size gap $\delta_\text{numerical},v,$ and $R$:
		\bea
		\delta_\text{numerical} = \frac{4v}{3R} + {\cal O}(\frac{1}{R^2}).
		\eea
		The different linear coefficients of $\delta_\text{numerical}$ and $\delta_\text{analytical}$ comes from the geometric difference between a cylinder and a hexagonal prism. Therefore, the validity of the analytical boundary theory for the DSM model is further confirmed by the numerical results of the finite-size gap in Fig. \ref{Fig: Finite size effect}. 
		
	\section{Real-space Representation for the DSM Hamiltonian}
	We now provide a real-space representation for the Hamiltonian described in Eq. \ref{Eq: DSM tight binding} in the second-quantization language.
	\begin{widetext}
		\begin{eqnarray}
		H &=& t\sum_{\langle {\bf r},{\bf r}'\rangle,s,\sigma} c^{\dagger}_{{\bf r}',s,\sigma} c_{{\bf r},s,\sigma} + t_z \sum_{{\bf r},s,\sigma} c^{\dagger}_{{\bf r},s,\sigma} c_{{\bf r+a}_z,s,\sigma} - \mu \sum_{{\bf r},s,\sigma} c^{\dagger}_{{\bf r},s,\sigma} c_{{\bf r},s,\sigma} -i v_0 \sum_{\langle {\bf r},{\bf r}'\rangle,s,\sigma\neq \sigma'} e^{-i s \theta_{{\bf r},{\bf r}'}} c^{\dagger}_{{\bf r}',s,\sigma'} c_{{\bf r},s,\sigma} \nonumber \\
		&& -\frac{iu_z}{2} \sum_{\langle {\bf r},{\bf r}'\rangle,s\neq s',\sigma\neq \sigma'} e^{-i2\theta_{{\bf r},{\bf r}'}} [c^{\dagger}_{{\bf r'+a}_z,s',\sigma'}c^{\dagger}_{{\bf r},s,\sigma} - c^{\dagger}_{{\bf r'-a}_z,s',\sigma'}c^{\dagger}_{{\bf r},s,\sigma}] + \text{h.c.},
		\end{eqnarray} 
	\end{widetext}
where $\langle {\bf r,r'}\rangle$ denotes the nearest neighboring atom positions within the horizontal plane. The phase angle characterizes the relative angle between the displacement vector for the hopping process and ${\bf a}_1$,
\begin{equation}
	\theta_{{\bf r},{\bf r}'} = \arccos \frac{({\bf r'-r})\cdot {\bf a}_1}{|{\bf r'-r}|}.
\end{equation}
Here $s=\uparrow/\downarrow$ and $\sigma=s,p$ are the spin and orbital indices. In the phase factor of the $v_0$-term, $s$ takes the value $\pm 1$ for $\uparrow$ and $\downarrow$, respectively. The chemical potential is $\mu=3t+t_z\cos k_0$, in terms of the parameters in Eq. \ref{Eq: DSM tight binding}. Physically, both $v_0$ and $u_z$ terms are generated from the spin-orbital coupling effect.  
	
\bibliography{HOTDSC}

\begin{thebibliography}{76}
\bibitem [{C3()}]{C3}%
  \BibitemOpen
  \href@noop {} {}\bibinfo {note} {Note that such 2d inversion symmetry is
  absent in a 3d $C_3$-symmetric system. Nevertheless, we have shown that $C_3$
  symmetry cannot protect the HOTDSC phase. Therefore, it is always reasonable
  to assume the existence of such 2d inversion symmetry in our target systems
  with $C_6$ symmetry}\BibitemShut {NoStop}%
\bibitem [{fot()}]{fote1}%
  \BibitemOpen
  \href@noop {} {}\bibinfo {note} {Note that the predicted gap in Eq. \ref{Eq:
  Surface gap closing condition} differs from the actual surface gap calculated
  in our lattice simulation by a constant factor $A=|\int_0^R dr f(r)^*
  \partial_r^2 f(r)|$. This factor comes from the spatial part $f(r)$ of our
  ansatz wavefunction in Eq. \ref{Eq: ansatz wavefunction}, which we have
  droped for simplicity in Eq. \ref{Eq: Fermi arc information}}\BibitemShut
  {NoStop}%
\end{thebibliography}
	
\end{document}